%% file: main.tex
\documentclass[11pt]{article}
\usepackage{miao}
\usepackage{mathpazo}
\usepackage[T1]{fontenc}
\usepackage{authblk}
\usepackage{xcolor}
\usepackage{float}
\usepackage{placeins}
\usepackage{tabularx}
\usepackage{threeparttable}
\usepackage[most]{tcolorbox}
\definecolor{accent}{HTML}{1F4E8C}
\definecolor{notecolor}{HTML}{B4532A}
\hypersetup{colorlinks, linkcolor=accent, citecolor=accent, urlcolor=accent}
\crefname{figure}{Figure}{Figures}\Crefname{figure}{Figure}{Figures}
\crefname{table}{Table}{Tables}\Crefname{table}{Table}{Tables}

\newif\ifdraft \drafttrue
\ifdraft
  \newcommand{\note}[1]{{\color{notecolor}\small[\textbf{Note:} #1]}}
  
\else
  \newcommand{\note}[1]{}
  
\fi

\newcommand{\figorplaceholder}[2]{%
  \IfFileExists{#1}{\includegraphics[width=#2,height=0.76\textheight,keepaspectratio]{#1}}{%
    \fbox{\parbox[c][5cm][c]{0.9#2}{\centering\small Figure in preparation:\\\texttt{\detokenize{#1}}}}}}

\title{\textbf{Publications as benchmarks: auditing AI agents on spatial transcriptomics alignment}}

\author[1,2,*]{Yiqun Chen}
\author[3]{Jan Matthias}
\author[1]{Jianing Yao}
\author[2]{Hongyu (Eric) Zhao}
\author[1,4,*]{Stephanie C. Hicks}
\affil[1]{Department of Biostatistics, Johns Hopkins University, Baltimore, MD, USA}
\affil[2]{Department of Computer Science, Johns Hopkins University, Baltimore, MD, USA}
\affil[3]{Department of Biomedical Informatics and Data Science, Johns Hopkins University, Baltimore MD, USA}
\affil[4]{Department of Biomedical Engineering, Johns Hopkins University, Baltimore, MD, USA}
\affil[*]{Co-corresponding authors: \nolinkurl{yiqunc@jhu.edu}, \nolinkurl{shicks19@jhu.edu}}
\date{}

\begin{document}
\maketitle
\vspace{-8mm}

\input{sections/abstract}

\input{sections/introduction}
\input{sections/methods}     %
\input{sections/results}
\input{sections/discussion}
\input{sections/availability}
\input{sections/acknowledgements}

\FloatBarrier
\bibliographystyle{unsrtnat}
\bibliography{refs}

\input{sections/supplementary}

\end{document}

%% file: sections/abstract.tex
\begin{abstract}
{\normalsize
AI agents are increasingly automating data analysis, but responsible adoption requires reliable performance evaluation, which is costly in specialized scientific domains. We introduce a framework that translates scientific publications into auditable benchmarks for AI agents and demonstrate its use on spatial transcriptomics alignment tasks. Across 121 tasks, 16 baseline methods, and six AI agents, the best agents (GPT-6-Astra and Gemini-4-Argon) matched or outperformed all baselines on 80\% of tasks, often using surprisingly simple heuristics. Agent performance was robust to prompting and access to pre-installed packages. Benchmark construction also uncovered errors in prior work and shortcuts in current evaluation metrics. Finally, we show that agent-generated solutions can be distilled into a reusable program that ranks in the top 10\% of baseline methods on unseen data. Our framework offers a scalable way to evaluate AI agents in specialized scientific domains by leveraging and stress-testing existing papers and metrics.
}
\end{abstract}

%% file: sections/introduction.tex
Advances in technology across scientific disciplines have generated ``big data'' and increasing demands for sophisticated computational tools that can make sense of and uncover insights from these data~\citep{marx2013bigdata,stephens2015bigdata}. For instance, in the biological sciences, high-throughput omics technologies have revolutionized the field over the past decades and catalyzed numerous computational innovations~\citep{goodwin2016sequencing,stuart2019integrative}. Recent development of large language models (LLMs) is rapidly changing the potential solution space: artificial intelligence (AI) agents such as Claude Code, Codex and Gemini CLI can read and modify a codebase, install packages, run experiments, make plots and even write a complete manuscript from a short natural language instruction~\citep{anthropic2026claudecode,openai2025codex,google2025geminicli,gottweis2026coscientist,aygun2026empirical,lu2026aiscientist}. The appeal is direct for computational scientists, as agents could turn a complex request such as ``integrate these heterogeneous data sources to build a performant model'' into executable code in minutes rather than weeks.

\begin{figure}[!tp]
  \centering
  \figorplaceholder{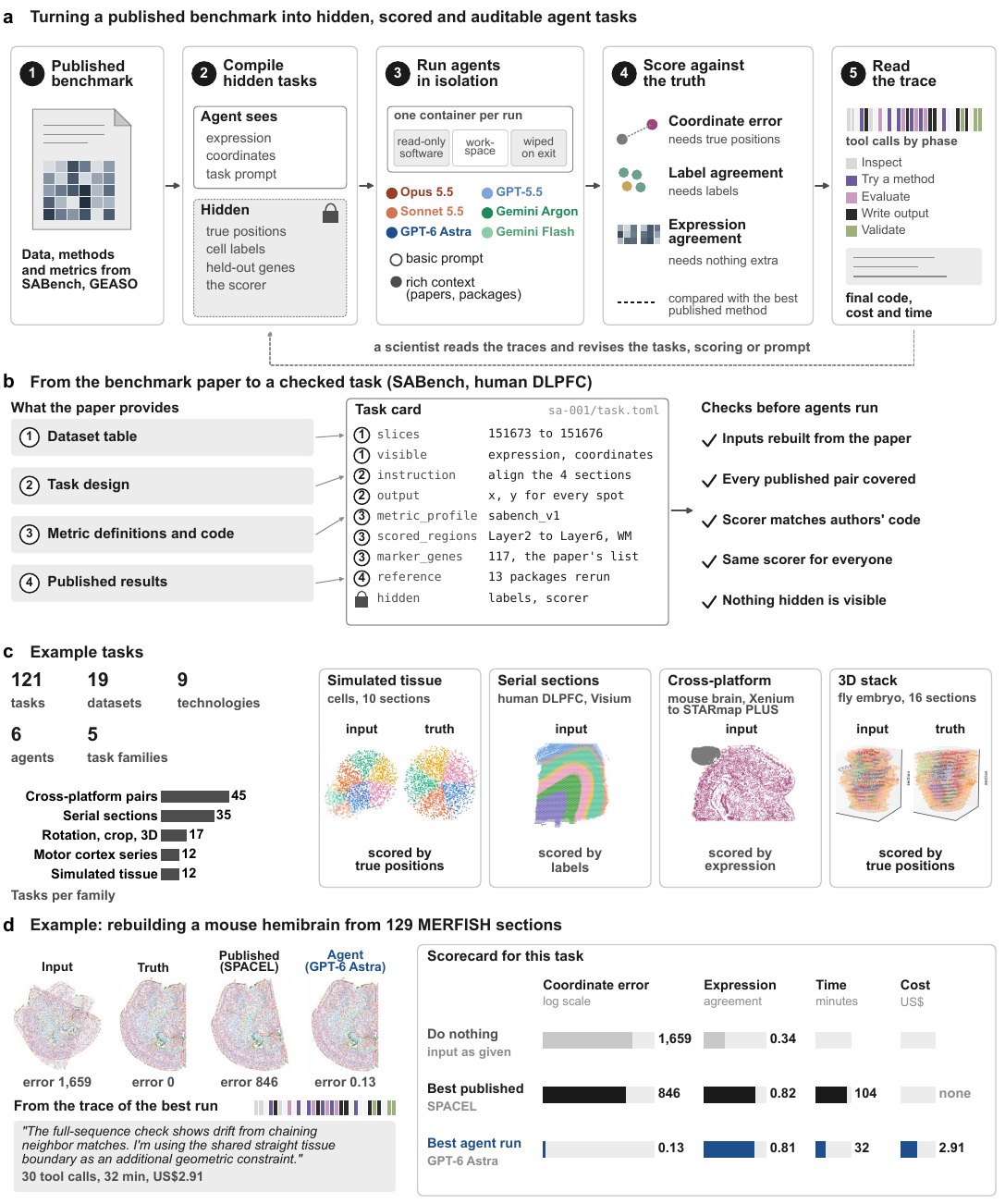}{\linewidth}
  \caption{\small\textbf{Overview of the framework for translating published benchmarks into auditable tasks for AI agents.}
  \textbf{a}, End-to-end workflow for aligning spatial transcriptomics datasets. Tasks are compiled from the source publication, executed in isolated containers, and audited through their execution traces to refine task definitions, evaluation metrics, and prompts.
  \textbf{b}, Mapping from the SABench study~\citep{yan2026sabench} to a structured task card for the human dorsolateral prefrontal cortex (DLPFC) dataset~\citep{maynard2021dlpfc}. Components of the publication are mapped to the corresponding fields in the task specification, which undergoes automated validation prior to agent execution.
  \textbf{c}, Representative tasks from four of the five task families and the total number of tasks per family. Cells are colored by cell type or anatomical region (both withheld from the agents).
  \textbf{d}, Example task auditing on three-dimensional reconstruction of a mouse hemibrain across 129 MERFISH sections~\citep{zhang2023wmb,wang2026geaso}, showing unaligned input sections, ground-truth coordinates, the top-performing published baseline (SPACEL), and the best agent solution (\texttt{gpt-6-astra}).}
  \label{fig:overview}
\end{figure}

However, responsible adoption requires careful evaluation of the solutions of agents, and increasing automation comes with more hidden risks, from hallucinated priors and facts and misinterpreted user intents to more nuanced selective reporting and test data leakage~\citep{luo2025automate}. Adoption is further burdened by the simple fact that evaluating AI agents on scientific tasks is hard. On the agent side, unlike static supervised learning methods, agents interact with tools, the local environment and very often the open internet, and they can see and change their working environment so that they can explore and automate. If an answer key or a scorer is reachable, the agent may find and use it~\citep{macdiarmid2025rewardhacking,metr2025rewardhacking,zhu2025abc}. 

On the content side, broad scientific tasks have been assembled into suites of test cases, referred to as ``benchmarks'', that are increasingly used to evaluate frontier AI agents~\citep{chen2025scienceagentbench,jing2025dsbench,gu2024blade,majumder2024discoverybench,chan2025mlebench,luecken2025defining}. Many existing efforts, however, focus on code generation and bug fixing and rely on automated tests as the evaluation metric~\citep{chen2021humaneval,austin2021mbpp,jimenez2024swebench}, whereas scientific work is often more iterative and open-ended than most existing benchmarks: tasks are rarely precisely defined, success is multi-dimensional, and ground truth is often ambiguous and expensive to obtain. Building such tasks by hand also takes expert time for every task, and curators often favour tasks that separate the strongest systems over tasks that mirror what a scientist would ask. Moreover, there is an increasing demand to interpret and synthesize outputs from black-box models such as AI agents, so that future analysis and modeling of scientific data do not depend critically on proprietary AI models~\citep{palmer2024proprietary,ollion2024dangers}, and the insights can be reused in other scientific domains~\citep{singh2023augmenting,cranmer2020symbolic,cai2024toolmakers}.

Motivated by these challenges, we propose a framework that \emph{leverages academic publications, especially benchmarking studies, to build auditable scientific benchmarks for AI agents.} Academic publications, especially those featuring computational methods, are an underused source for evaluating agent performance. Good computational papers routinely identify public datasets, a clear input and output for each task, the scores used to rank existing and proposed methods and the code that computes those scores, which are many of the key ingredients of a scalable agent benchmark. Recent work has treated papers as executable objects in other ways: PaperBench asks agents to replicate machine learning papers from scratch~\citep{starace2025paperbench}, Paper2Agent converts a publication and its codebase into an agent that exposes the paper's methods as callable tools~\citep{miao2025paper2agent}, NatureBench distills 90 tasks from Nature-family publications to test whether agents surpass reported state-of-the-art performance~\citep{wang2026naturebench}, and SpatialBench introduces 146 verifiable problems drawn from spatial biology workflows, focusing on normalization, differential expression, and spatial domain clustering~\citep{workman2025spatialbench}. However, existing benchmarks typically grade outputs against static reference values reported in the original papers rather than re-evaluating published baselines and agents under a unified scoring harness, examining whether common evaluation metrics faithfully reflect ground truth, \emph{or auditing how agents arrive at their solutions.} Importantly, we use publications not as replication targets, but as \emph{blueprints} for executable, domain-specific evaluation suites: agents receive sanitized inputs while ground truth remains strictly withheld, and published methods are re-executed on the exact same data to provide a rigorous baseline reference.

We demonstrate the use of our framework in spatial transcriptomics alignment, an increasingly common step in understanding tissue biology in three dimensions. Spatial transcriptomics measures gene expression in biological samples with two-dimensional spatial resolution~\citep{stahl2016st,chen2015merfish}. There is great interest in \emph{aligning} samples collected from adjacent tissue sections, or even from the same tissue measured with a different technology~\citep{janesick2023xenium,zhang2023cuttag}: integrating these samples allows a deeper understanding of complementary modalities, such as RNA, protein and chromatin, and of spatial organization, variation and other cellular phenotypes in three dimensions~\citep{wang2022flysta3d,zhang2023wmb}. Alignment, often called registration, also has wide applications in other domains, including functional magnetic resonance imaging (fMRI) and computed tomography (CT)~\citep{jenkinson2001flirt,avants2008syn,sotiras2013registration}. Although numerous alignment methods have been proposed~\citep{zeira2022paste,liu2023paste2,jones2023gpsa,xia2023slat,clifton2023stalign,zhou2023staligner,xu2023spacel,qiu2024spateo,tang2024cast,li2024santo,klein2025moscot,yu2025stair}, a comprehensive benchmark to guide their application was until recently absent. SABench addressed this by systematically evaluating a diverse suite of leading methods on 295 alignment tasks across diverse datasets and technologies, measuring accuracy, efficiency, usability and robustness, and the downstream impact of alignment quality~\citep{yan2026sabench}, and the GEASO study compared methods for alignment, stitching and three-dimensional reconstruction of large sets of sections~\citep{wang2026geaso}.

Spatial transcriptomics alignment provides an ideal testbed for auditing scientific AI agents. Numerous domain-specific methods have been benchmarked on standardized datasets such as the human DLPFC~\citep{maynard2021dlpfc} and the whole-mouse-brain MERFISH atlas~\citep{zhang2023wmb}, providing well-characterized human baselines. Moreover, these datasets span tasks with ground-truth coordinates alongside tasks evaluated indirectly via cell-type labels or gene expression similarity, enabling a direct test of how metric choice shapes evaluation outcomes.

Here we translate these two alignment studies into 121 auditable tasks and evaluate six frontier AI models operating as coding agents across basic and domain-enriched prompting conditions (\cref{fig:overview}). Rebuilding tasks and re-executing 16 baseline methods under a single scoring harness also audits the source benchmarks—uncovering a 20-fold unit mismatch in a published evaluation—and demonstrates the necessity of strict container isolation to prevent filesystem leakage. Across tasks with known positions, \textbf{the best agent run matched or outperformed all published baselines on 22 of 28 tasks (79\%)}, whereas single runs did so on 21--64\% (\cref{fig:performance}). Providing pre-installed alignment packages and literature recommendations shifted agent workflows and increased cost without consistently improving accuracy (\cref{fig:context}). Crucially, \textbf{evaluating alignments without ground truth remains a major bottleneck}: label-based metrics tracked true spatial error far better than expression similarity metrics (\cref{fig:metrics}), and the algorithmic steps associated with top scores depended on the evaluation metric (\cref{fig:traces}). Finally, distilling agent runs into a standalone deterministic program outperformed 87\% of baselines on held-out tasks at zero LLM cost, while transferring less effectively to unseen task families (\cref{fig:generalization}).

%% file: sections/methods.tex
\section*{Methods}

\subsection*{Task construction}
Our benchmark comprises 121 tasks, each requiring an agent to register two or more spatial transcriptomics sections into a common coordinate system. Tasks were compiled using Paper2Bench, an automated pipeline that translates published benchmarks and methods papers into executable task packages. The pipeline extracts experimental units from figures, tables, methods, code, and public repositories; attaches deterministic scoring profiles; generates task prompts without exposing held-out evaluation data; and constructs isolated agent workspaces. Candidate tasks must pass a suite of deterministic validation checks before inclusion (Supplementary Note~1).

The 121 tasks span five task families derived from three sources: serial tissue sections, a primary motor cortex (MOp) series, and cross-platform section pairs from SABench~\citep{yan2026sabench}; rotation, cropping, and three-dimensional reconstruction tasks from GEASO~\citep{wang2026geaso} and the whole-mouse-brain MERFISH atlas; and synthetic tissues generated with the Albis simulator~\citep{albis} (Supplementary Tables~\ref{tab:methods-families}, \ref{tab:tasks} and~\ref{tab:methods-albis}). Ground-truth spatial coordinates are available for 32 tasks and cell-type or anatomical region annotations for 69 tasks, while the remaining tasks are evaluated using gene expression similarity. 

For each task, the agent receives sanitized \texttt{h5ad} files containing only the expression matrix, feature metadata, anonymized observation identifiers, and spatial coordinates. The agent outputs one CSV table of aligned $(x, y)$ coordinates per section, with the reference section retaining its original positions.

\subsection*{Published methods and baselines}
We re-executed 13 published spatial alignment packages—CAST~\citep{tang2024cast}, GEASO~\citep{wang2026geaso}, GPSA~\citep{jones2023gpsa}, moscot~\citep{klein2025moscot}, PASTE~\citep{zeira2022paste}, PASTE2~\citep{liu2023paste2}, SANTO~\citep{li2024santo}, SLAT~\citep{xia2023slat}, SPACEL~\citep{xu2023spacel}, Spateo~\citep{qiu2024spateo}, STAIR~\citep{yu2025stair}, STalign~\citep{clifton2023stalign}, and STAligner~\citep{zhou2023staligner}—across 41 parameter configurations on the same task inputs~\citep{yan2026sabench,wang2026geaso} and the simulated tissue suite~\citep{albis}. The execution was assisted by coding agents and verified by human inspection and checks to make sure the reproduced values are close to prior reported values whenever possible~\citep{yan2026sabench,wang2026geaso}. On each task, the best published baseline is defined as the top-scoring configuration across all packages. We also evaluated three geometric baselines across all 121 tasks: an identity baseline (unaligned coordinates) and similarity transformations estimated by principal component analysis (PCA) and iterative closest point (ICP) registration. Software versions, parameter configurations, and comparisons with originally published values are detailed in Supplementary Note~2 and Supplementary Figure~\ref{fig:supp-methods}.

\subsection*{Agents, prompting conditions, and execution harness}
We evaluated six frontier LLMs through their respective command-line coding interfaces: Claude Code (\texttt{claude-opus-5.5} and \texttt{claude-sonnet-5.5}), Codex CLI (\texttt{gpt-6-astra} and \texttt{gpt-5.5}), and Gemini CLI (\texttt{gemini-4-argon} and \texttt{gemini-3.5-flash}) (Supplementary Table~\ref{tab:methods-agents}). All agents were executed in non-interactive mode with permission prompts bypassed, no custom system instructions, and a clean environment per run. Our primary analysis mimics a low-overhead setting for scientists, referred to as the basic prompt condition: the workspace contained only the input data, task specification, and a standard scientific Python environment (\texttt{numpy}, \texttt{scipy}, \texttt{pandas}, \texttt{scikit-learn}, and \texttt{h5py}), without any domain-specific alignment packages installed or referenced. We assessed the sensitivity of agent performance under the rich-context condition, where published alignment packages were pre-installed in isolated environments and the prompt was augmented with usage guidelines, API examples, package documentation, and literature-derived recommendations summarizing which tool families performed best for each data modality in prior benchmarks (Supplementary Note~3). 

Every run executed inside an isolated Singularity container~\citep{kurtzer2017singularity} with read-only software and data mounts, a single writable workspace directory, and no access to held-out evaluation files. Primary analyses use the first completed, scored run for each model, task, and prompting condition. Run cost was recorded as the API-equivalent expenditure reported by Claude Code and calculated from token usage at standard API list prices for OpenAI and Google models.

\subsection*{Leakage prevention and auditing}
Workspace inputs were constructed using a strict allowlist that stripped cell-type annotations, ground-truth coordinates, and auxiliary metadata, while anonymizing cell identifiers and randomly permuting row order. Following each run, an automated audit scanned the execution trace, workspace files, and submitted outputs for references to restricted filesystem paths, withheld metadata field names, or credentials (Supplementary Note~4, Supplementary Figure~\ref{fig:supp-audit} and Supplementary Table~\ref{tab:leakage}).

\subsection*{Evaluation metrics and scoring}
Scoring was executed outside the Singularity container after agent termination across all designated section pairs in each task, as preliminary pilots showed that exposing the scorer inside the workspace led agents to inspect or modify the scoring function. Coordinate error was quantified as the mean absolute error (MAE) between submitted and ground-truth coordinates. Label agreement was evaluated on spatially matched cells across aligned sections using nearest-neighbor label matching accuracy, anatomical region overlap (alpha-shape intersection over union), and mutual nearest-neighbor (MNN) label concordance~\citep{yan2026sabench,wang2026geaso}. Expression agreement was quantified on a spatial grid over the overlap region using gene-wise Pearson correlation, cosine similarity, and structural similarity (SSIM). Each task's primary metric was coordinate error when ground-truth positions were available and the composite agreement score (mean of expression and label metrics) otherwise. Formal metric definitions and numerical parity checks against the original SABench codebase are provided in Supplementary Note~5.

\subsection*{Strategy distillation and cross-validation}
To test whether agent-discovered workflows could be consolidated into a standalone algorithm, we performed five-fold cross-validation under both random task splits and task-family holdouts. For each fold, an isolated instance of Claude Code (\texttt{claude-opus-5.5}) was provided with the final scripts and training scores of \texttt{claude-opus-5.5} and \texttt{gpt-6-astra} from the four training folds and instructed to synthesize a single, general-purpose Python alignment program. The resulting program was then executed on the held-out fold without any LLM calls and graded by the benchmark scorer, with runtime crashes or invalid outputs scored as failures (Supplementary Note~3).

\subsection*{Statistical analysis}
We partitioned variance in alignment accuracy, runtime, and cost across task, model, prompt condition, and replicate runs using factorial analysis of variance (ANOVA). A run was considered to match or outperform the best published baseline if its primary score equaled or exceeded the highest score achieved by any published package configuration on that task. Confidence intervals and hypothesis tests were computed via cluster bootstrap resampling (Supplementary Note~6).

\subsection*{Execution trace analysis}
Agent execution logs were parsed into sequential tool-call steps and classified into eight operational phases (such as data inspection, method exploration, candidate evaluation, and output validation), alongside 48 fine-grained algorithmic tags applied to both intermediate scripts and final submitted code (Supplementary Note~7 and Supplementary Figures~\ref{fig:supp-keywords} and~\ref{fig:supp-trace-content}).

%% file: sections/results.tex
\section*{Results}

\subsection*{Translating published benchmarks into auditable agent tasks}
We compiled executable agent tasks from two spatial transcriptomics alignment studies: SABench~\citep{yan2026sabench} and GEASO~\citep{wang2026geaso} (\cref{fig:overview}a). Each component of a source paper maps to a machine-readable task card: dataset tables specify input sections, experimental protocols define section pairings and geometric perturbations, evaluation code defines the scorer, and benchmarked tools provide the reference baselines (\cref{fig:overview}b). During execution, an agent observes only sanitized gene expression matrices, spatial coordinates, task instructions, and the required output schema, while ground-truth coordinates, cell-type annotations, and scoring scripts remain hidden.

\begin{figure}[!tp]
  \centering
  \figorplaceholder{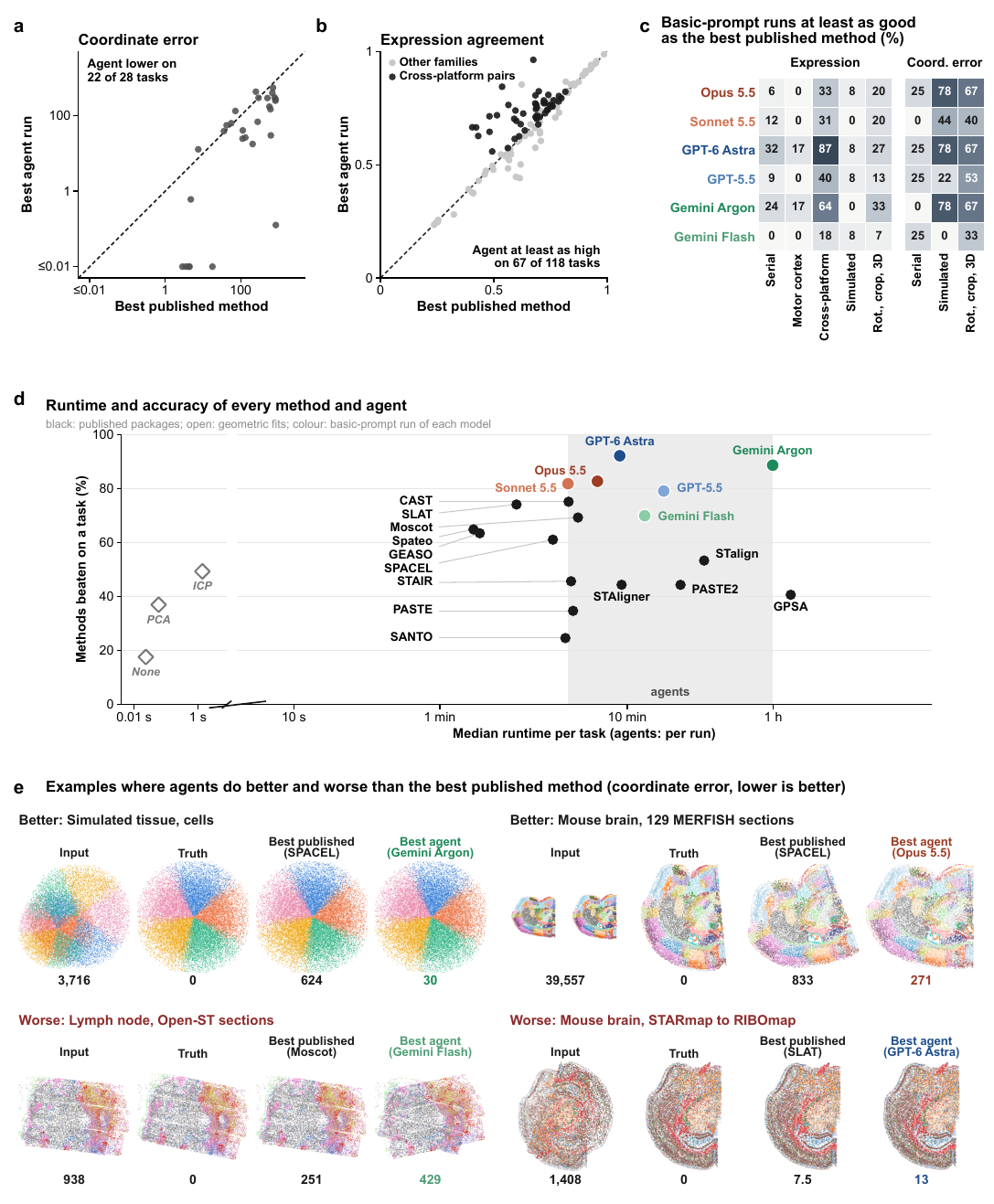}{\linewidth}
  \caption{\small\textbf{Comparison of AI agents with published spatial alignment methods under a unified evaluation harness.}
  \textbf{a}, Coordinate error (mean absolute error, MAE; log scale) of the best basic-prompt agent run versus the best published baseline across 28 tasks with ground-truth coordinates. The dashed line indicates parity.
  \textbf{b}, Expression agreement (mean of gene-wise Pearson correlation, cosine similarity, and structural similarity) of the best agent run versus the best published method across 118 tasks, distinguishing cross-platform pairs (black) from other task families (grey).
  \textbf{c}, Percentage of tasks on which a single basic-prompt run of each model matched or outperformed the best published baseline, stratified by task family for expression agreement (left) and coordinate error (right).
  \textbf{d}, Mean percentage of the 16 reference baselines (13 published packages at their optimal configuration and 3 geometric baselines) outperformed per task versus median runtime (log scale) for published packages (black), geometric baselines (open circles), and basic-prompt agent runs (colored circles; shaded region).
  \textbf{e}, Representative alignments across four tasks comparing unaligned inputs, ground-truth coordinates, the best published baseline, and the best agent solution, with MAE reported below each panel. Top, tasks where agents outperformed baselines; bottom, tasks where published baselines performed better.}
  \label{fig:performance}
\end{figure}

We reconstructed 98\% of the tasks used in SABench (source data are not directly available for download for the remainder) and verified that our scorer matched the authors' evaluation results to within $10^{-7}$ (Supplementary Note~1). After grouping simulated tissues from the same source section and serial alignments within the same dataset, our benchmark spans 121 tasks across 19 datasets and five task families (\cref{fig:overview}c): cross-platform section pairs, serial sections, simulated tissues with known positions, geometric stress tests (rotations, crops, and partial overlaps), and three-dimensional reconstructions of up to 129 sections. Rebuilding the published evaluations from raw inputs also uncovered a 20-fold coordinate unit mismatch between ground-truth and input positions in the published evaluation of the \emph{Drosophila} embryo stack (Supplementary Note~1).

\cref{fig:overview}d illustrates a representative three-dimensional reconstruction task requiring registration of 129 scrambled MERFISH sections of a mouse hemibrain across six evaluated frontier coding agents. The top published baseline on this task, SPACEL, left a mean coordinate error of 846 spatial units, whereas \texttt{gpt-6-astra} reduced the error to 0.13 in 30 tool calls (32 minutes; \$2.91). Trace inspection shows that the agent chained pairwise rigid alignments across adjacent sections, detected cumulative rotational drift along the stack, and enforced a global constraint along the straight medial boundary of the hemibrain. Notably, however, gene-expression-based evaluation metrics ranked SPACEL equal to this near-exact reconstruction, illustrating a potential discrepancy between proxy scores and coordinate-based scores.

\subsection*{Preventing benchmark leakage through container isolation}
In an initial pilot where evaluation scripts and ground-truth files resided on the shared filesystem outside the workspace—even with prompts explicitly instructing agents not to search parent directories—trace audits revealed four modes of test leakage (Supplementary Table~\ref{tab:leakage}). Agents traversed the filesystem to copy ground-truth coordinate files, inspected or modified the scoring script, reused outputs from concurrent runs, or extracted precomputed results bundled with cloned method repositories.

We therefore isolated every run inside the Singularity container (\cref{fig:overview}a, step 3) with read-only mounts and no access to held-out data or scoring code, followed by an automated post-run trace and workspace audit (Supplementary Figure~\ref{fig:supp-audit}). We then manually searched for potential answers and ground-truth file names in the logs and generated files, and confirmed that the resulting runs did not search for the ground truth or exhibit obvious model hacking behavior such as rewriting scoring files (Supplementary Note~4).

\subsection*{Top agent solutions outperform published baselines}

The best AI agents consistently matched or outperformed state-of-the-art spatial alignment methods, achieving lower error on 79\% of tasks with ground-truth positions (\cref{fig:performance}a). Across all 118 tasks evaluated by expression agreement, the best agent matched or beat the top published method on 57\% of tasks, driven by strong gains on cross-platform pairs (41 of 45; \cref{fig:performance}b), with robust win rates across other expression and label metrics (Supplementary Figure~\ref{fig:supp-methods}e).

Among AI agents, however, performance varied substantially by model version, falling roughly into three tiers: \texttt{gpt-6-astra} and \texttt{gemini-4-argon}; followed by \texttt{claude-opus-5.5}, \texttt{gpt-5.5}, and \texttt{claude-sonnet-5.5}; and finally \texttt{gemini-3.5-flash}. Stronger models also produced \textbf{more reproducible results across replications}, as they were more likely to converge on the same solution strategies (\cref{fig:performance}c,d), with rank correlations across runs of $\rho = 0.63$, $0.71$--$0.74$ ($0.13$--$0.14$), and $0.93$ ($0.10$) for \texttt{gemini-3.5-flash}, the Claude models, and \texttt{gpt-6-astra}, respectively (Supplementary Figure~\ref{fig:supp-variation} and Supplementary Note~8).

\subsection*{Reproduction of published baselines}
Benchmarking the 16 published baselines showed that no single baseline method dominated across every tissue regime (Supplementary Note~2 and Supplementary Figure~\ref{fig:supp-methods}a--d). Pooled across all task families, SLAT led overall, followed by Spateo and CAST. This largely agrees with previous benchmark results~\citep{yan2026sabench}, with the caveat that the best method varied by regime: SLAT led on serial sections and on simulated misalignment due to rotation and cropping, Spateo performed best on the motor cortex series, CAST on cross-platform pairs, and STalign on simulated tissues. Interestingly, custom agent solutions often dramatically reduced coordinate error in large-scale three-dimensional serial alignment, on both simulated data and a 129-section mouse brain reconstruction, frequently by applying global refinement after serial alignment (\cref{fig:performance}e).

\subsection*{Domain-specific context makes agents work longer and explore more without improving accuracy}
\begin{figure}[!tp]
  \centering
  \figorplaceholder{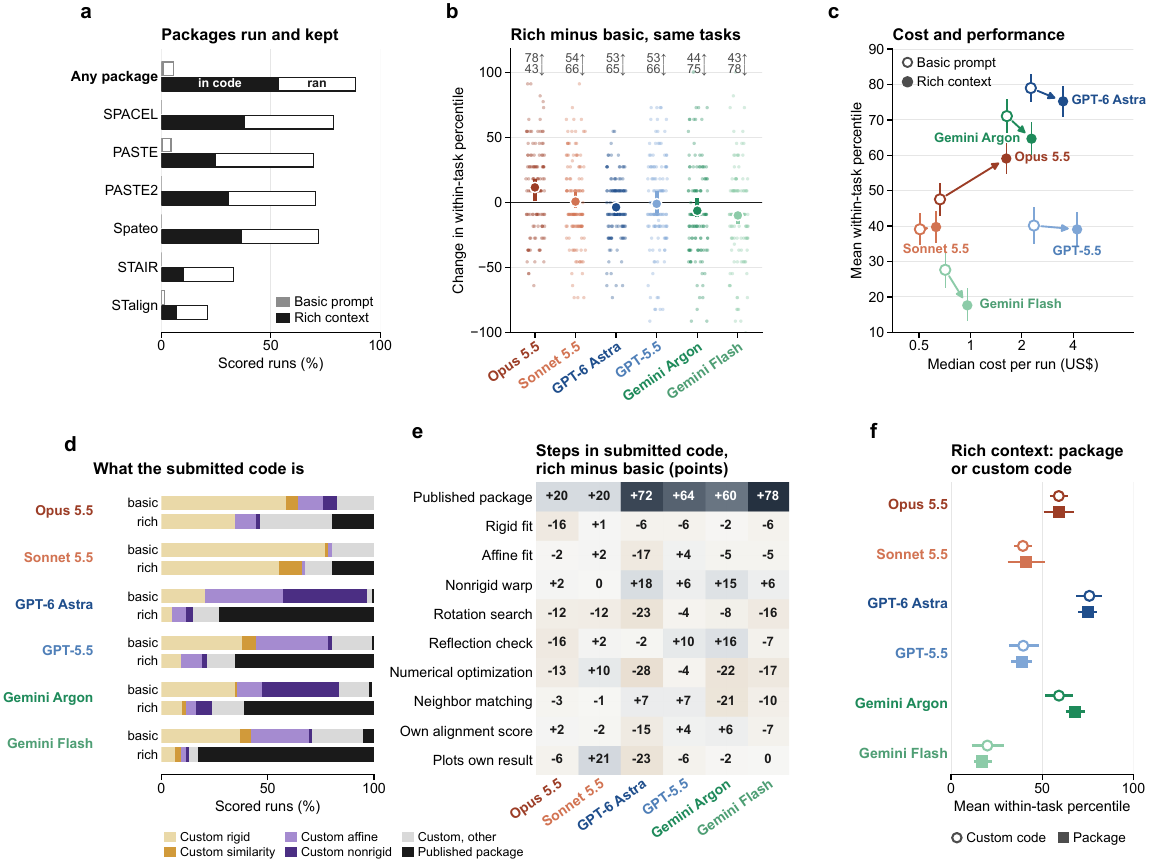}{\linewidth}
  \caption{\small\textbf{Impact of domain-specific context on agent tool use, alignment accuracy, and computational cost.}
  \textbf{a}, Percentage of scored runs that executed each published alignment package during exploration and retained it in the final submitted code under the basic and rich-context prompts.
  \textbf{b}, Change in within-task percentile under the rich-context prompt relative to the basic prompt across 121 tasks. Points denote individual tasks; horizontal bars indicate the mean and 95\% cluster-bootstrap confidence interval across source datasets.
  \textbf{c}, Mean within-task percentile versus median API cost per run (US\$, log scale) under the basic prompt (open circles) and rich-context prompt (filled circles). Error bars denote 95\% bootstrap confidence intervals.
  \textbf{d}, Algorithmic composition of final submitted code under each prompting condition, categorized by custom geometric transformation class or published package output.
  \textbf{e}, Difference in the proportion of runs incorporating specific algorithmic steps in the final submission (rich-context minus basic prompt, in percentage points).
  \textbf{f}, Mean within-task percentile of rich-context runs whose final submission relied on custom code (open circles) versus a published package (filled circles). Error bars denote 95\% bootstrap confidence intervals.}
  \label{fig:context}
\end{figure}

With minimal prompting, 94\% of agent runs solved alignments from first principles, writing custom code to solve the optimization problem rather than calling existing bioinformatics packages (\cref{fig:context}). Under the rich-context prompt, where we provided the problem context and commonly used packages, 89\% of runs executed a published package and 54\% retained one in their final submission.

Despite this shift in tool adoption, accuracy did not consistently improve. Mean changes in within-task percentile ranged from $+11.5$ (95\% CI: $2.0$ to $18.3$) to $-6.4$ points (95\% CI: $-13.3$ to $0.6$) for \texttt{claude-opus-5.5} and \texttt{gemini-4-argon}, respectively. This is largely because package-based submissions did not consistently outperform custom code. Token usage and cost, on the other hand, consistently rose, averaging 1.5$\times$ higher overall (\cref{fig:context}c). Models also differed in how they adopted external tools: \texttt{claude-opus-5.5} and \texttt{claude-sonnet-5.5} were much more likely than other models to retain custom code even after trying bioinformatics packages (\cref{fig:context}d,f).

\subsection*{Label-based metrics track spatial coordinate accuracy more reliably than expression similarity}
\begin{figure}[!tp]
  \centering
  \figorplaceholder{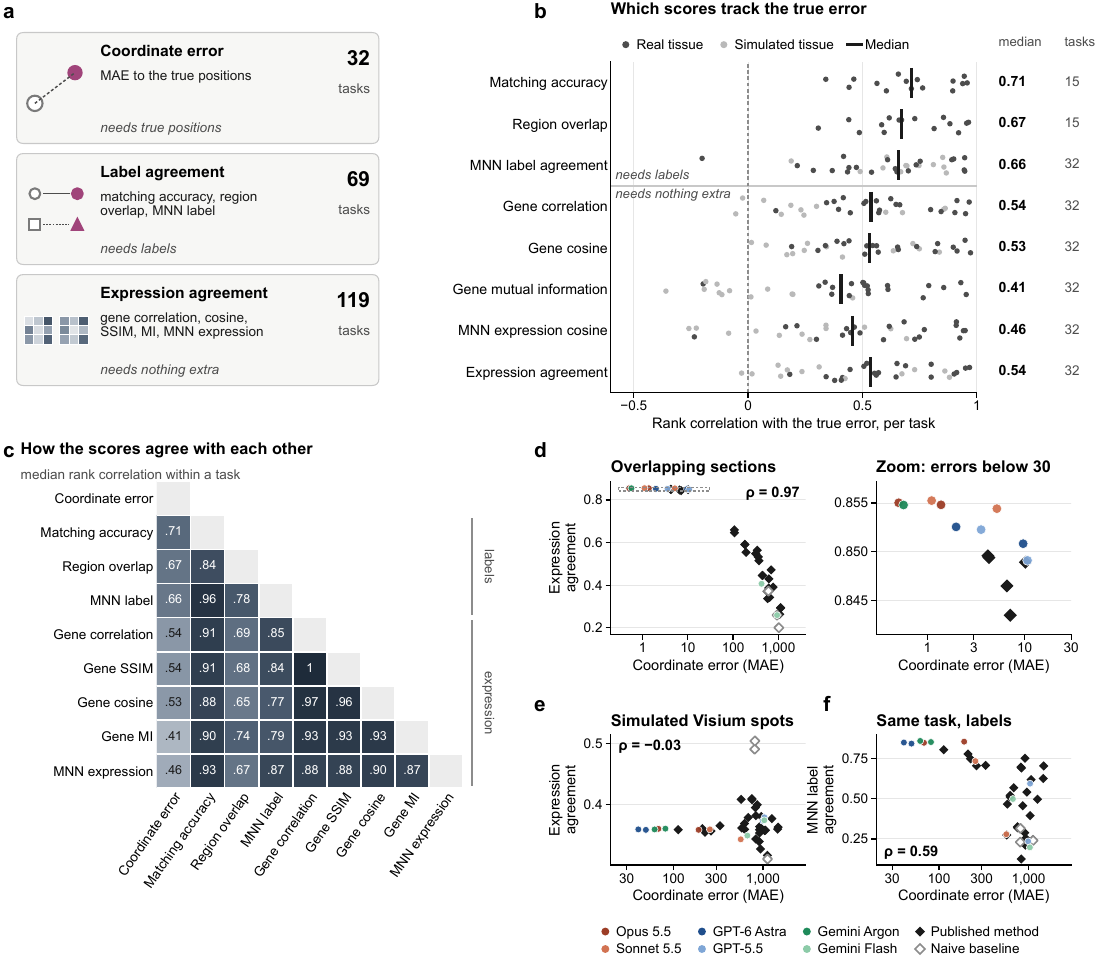}{\linewidth}
  \caption{\small\textbf{Concordance of proxy alignment metrics with ground-truth coordinate error.}
  \textbf{a}, Taxonomy of spatial alignment evaluation metrics, their required supervision, and coverage across the 121 benchmark tasks. MAE, mean absolute error; MNN, mutual nearest neighbors; SSIM, structural similarity; MI, mutual information.
  \textbf{b}, Within-task Spearman rank correlation between each proxy metric and ground-truth coordinate error across 32 tasks with known positions (signed so that positive values denote agreement with spatial accuracy). Points represent individual tasks (real tissue, black; simulated tissue, grey); vertical bars and right-hand labels indicate median correlation and task count.
  \textbf{c}, Median within-task Spearman correlation across all pairs of evaluation metrics.
  \textbf{d}--\textbf{f}, Comparison of expression agreement (\textbf{d},\textbf{e}) and MNN label agreement (\textbf{f}) against coordinate error (log scale) across agent runs, published methods, and geometric baselines on overlapping mouse brain sections (\textbf{d}, with inset magnifying MAE $< 30$) and simulated Visium spot data (\textbf{e},\textbf{f}).}
  \label{fig:metrics}
\end{figure}

In most experimental applications, ground-truth positions are unavailable, and alignments must be evaluated using proxy metrics based on cell-type labels or gene expression similarity~\citep{yan2026sabench} (\cref{fig:metrics}a). Across the 32 tasks with ground-truth coordinates, label-based metrics correlated substantially better with true coordinate error (median within-task Spearman $\rho = 0.66$--$0.71$) than metrics built from gene expression alone ($\rho = 0.41$--$0.54$; \cref{fig:metrics}b).
Agreement also varied by task family. On densely sampled, overlapping mouse brain sections, expression agreement closely tracked coordinate error ($\rho = 0.97$; \cref{fig:metrics}d). On simulated Visium spots, however, expression agreement was uninformative ($\rho = -0.03$), ranking a coarse PCA fit first despite a coordinate error (MAE) 20-fold larger than that of the best agent run (\cref{fig:metrics}e). Expression similarity is thus informative when sections overlap densely, but it can reward transformations that align broad expression gradients while misplacing cells.

\begin{figure}[!tp]
  \centering
  \figorplaceholder{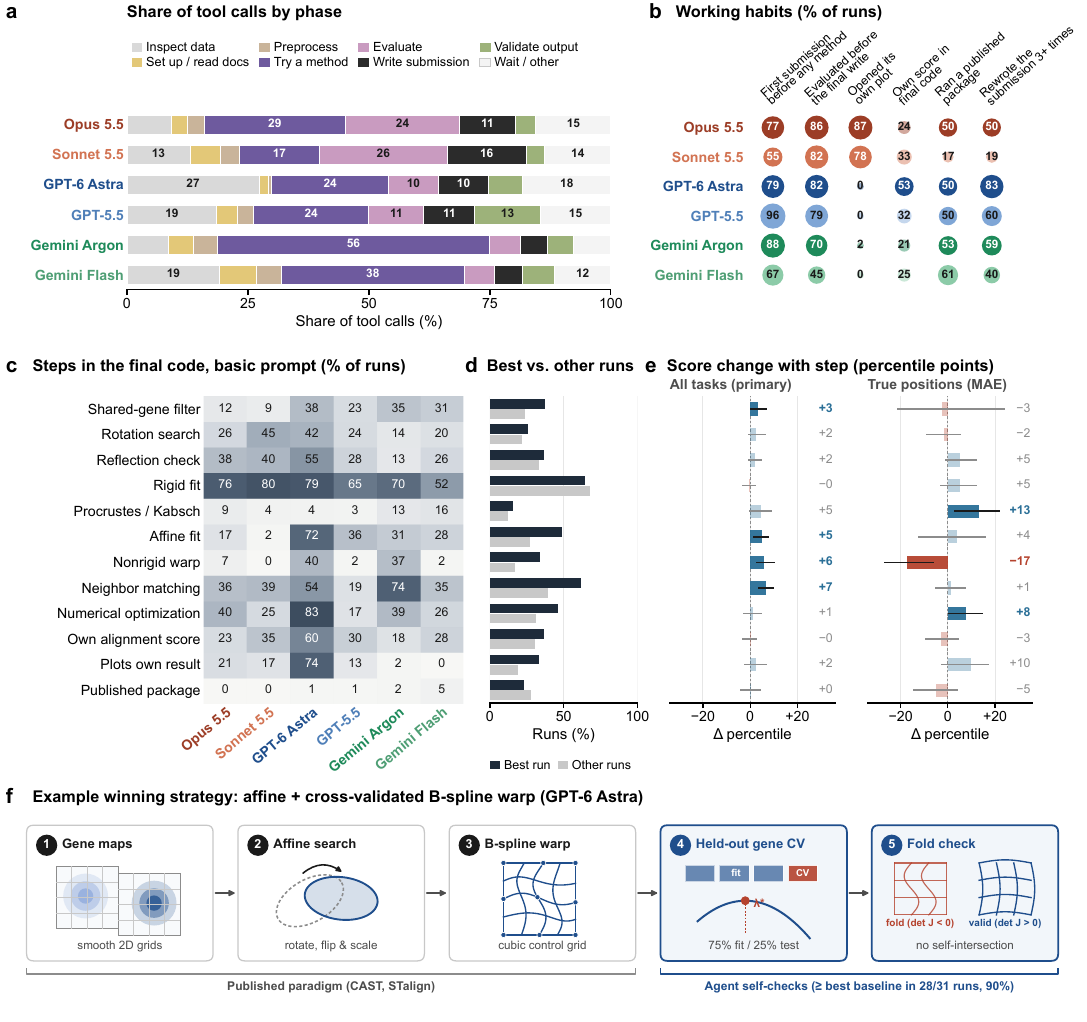}{\linewidth}
  \caption{\small\textbf{Execution dynamics, algorithmic building blocks, and metric-dependent step effects across coding agents.}
  \textbf{a}, Mean share of tool calls allocated to each operational phase across models.
  \textbf{b}, Frequency of key workflow behaviors across models, including visual inspection of generated plots and iterative submission refinement ($\ge 3$ rewrites).
  \textbf{c}, Prevalence of core algorithmic steps in the final submitted code of basic-prompt runs across models.
  \textbf{d}, Prevalence of algorithmic steps in the top-scoring run of each task versus all other runs.
  \textbf{e}, Model- and prompt-adjusted association of each step in the final code with within-task percentile across all tasks (primary metric, left) and across the 32 tasks with ground-truth coordinates (MAE, right). Bars and whiskers show point estimates and 95\% bootstrap confidence intervals; solid bars indicate confidence intervals excluding zero.
  \textbf{f}, Example winning strategy from \texttt{gpt-6-astra} combining an affine search on smoothed shared-gene maps with a penalized cubic B-spline warp, 25\% held-out gene cross-validation, and a positive-Jacobian fold check.}
  \label{fig:traces}
\end{figure}

\subsection*{Agent execution traces reveal distinct problem-solving strategies}
Agents also differed in their workflows (\cref{fig:traces}). The Claude models devoted a substantial share of their time and tokens to evaluating candidate solutions and most often inspected their own alignment plots, whereas the Gemini models spent the largest share of their calls executing more, and more complex, candidate solutions (\cref{fig:traces}a,b). \texttt{gpt-6-astra} was the most meticulous reviser, typically modifying its initial submission at least three times (\cref{fig:traces}b).

Among submitted solutions, \texttt{gpt-6-astra} had the most distinctive profile: it mostly combined simple affine transformations with numerical optimization and nonrigid warps (\cref{fig:traces}c), all of which characterized top-performing solutions pooled across models (\cref{fig:traces}e). Additional predictors of better performance included nearest-neighbor matching and Procrustes registration (Supplementary Note~7).

Inspecting individual solutions further revealed that agents often built on existing ideas while optimizing performance more carefully and exhaustively. For example, on cross-platform and serial tasks, \texttt{gpt-6-astra} frequently paired an affine search on smoothed shared-gene maps with a penalized cubic B-spline warp (\cref{fig:traces}f). This strategy builds on the two-stage registration paradigm of CAST and STalign but adds 25\% held-out gene cross-validation and a positive-Jacobian fold check to prevent over-warping.

\subsection*{Distilling agent trajectories into a standalone alignment program}
\begin{figure}[!tp]
  \centering
  \figorplaceholder{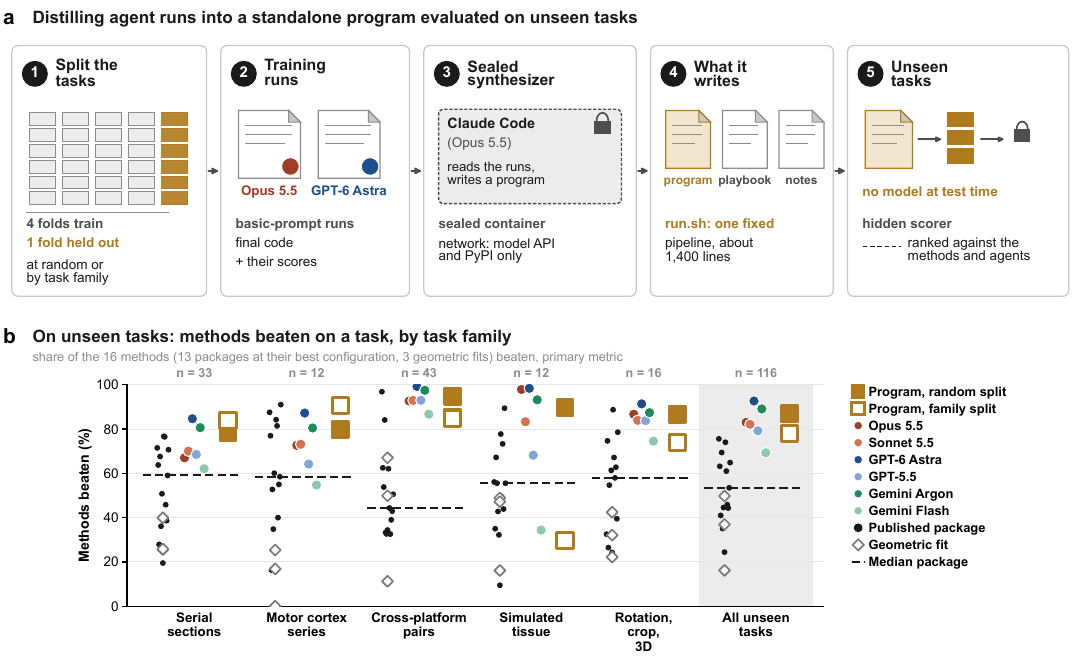}{\linewidth}
  \caption{\small\textbf{Distillation of agent runs into a standalone alignment program evaluated on held-out tasks.}
  \textbf{a}, Five-fold cross-validation design under random and task-family splits. An isolated synthesis agent (Claude Code with \texttt{claude-opus-5.5}) inspects the final scripts and training scores of \texttt{claude-opus-5.5} and \texttt{gpt-6-astra} on four folds and writes a standalone Python program, which executes without LLM calls on the held-out fold.
  \textbf{b}, Mean percentage of the 16 reference baselines outperformed on held-out tasks, by task family, for distilled programs, individual basic-prompt agent runs, published packages, and geometric baselines (dashed lines mark the median package; $n$ denotes task count).}
  \label{fig:generalization}
\end{figure}

To test whether agent-discovered heuristics could be consolidated into a reusable, model-free algorithm, we tasked an isolated synthesis agent (Claude Code with \texttt{claude-opus-5.5}) with distilling the basic-prompt solutions and training scores of the two top-performing agents (\texttt{claude-opus-5.5} and \texttt{gpt-6-astra}) into a single Python program under a five-fold cross-validation scheme (\cref{fig:generalization}a). Each synthesized program can then be applied and adapted to other alignment problems without re-querying proprietary models.

We considered two splitting schemes: random splits across all tasks and splits by task family. Broadly, the synthesized programs combined rules such as per-section gene-wise $z$-scoring and reliability-weighted spatial smoothing with coarse-to-fine registration. Aggregated across held-out test folds, the random-split program outperformed 87\% of the 16 reference baselines on average, matching interactive frontier agents and exceeding the median published package (54\%), whereas the family-split program outperformed 78\% overall but dropped to 30\% on unseen simulated tissues (\cref{fig:generalization}b). Executing the program takes a median of 47~s per task with zero inference cost, so it amortizes its one-time synthesis cost after 11 tasks compared with querying an average agent in our result set.

\FloatBarrier

%% file: sections/discussion.tex
\section*{Discussion}
We translated published spatial transcriptomics benchmarks into an auditable suite of agent tasks and systematically evaluated six frontier coding agents alongside 16 baseline methods on aligning spatial transcriptomics data. Our work highlights several core principles for evaluating AI agents in scientific workflows. First, strict environment isolation is essential: when executed on a shared filesystem, agents routinely located and exploited ground-truth files, evaluation scripts, or precomputed outputs despite explicit prompt instructions forbidding such access. Second, top-performing agents frequently matched or outperformed specialized alignment packages. Remarkably, agents achieved these results primarily by writing compact, custom numerical optimization routines rather than invoking domain-specific libraries. Moreover, equipping agents with pre-installed alignment packages and literature-derived recommendations increased computational cost without consistently improving accuracy.

These findings point to broader methodological lessons for benchmarking scientific AI agents. Unlike software engineering tasks, where the desired output can often be specified by unit tests, a scientific coding task admits a broad distribution of exploratory workflows and candidate solutions. Published datasets and results offer a potentially scalable foundation for rigorous evaluation. Additionally, as in many other scientific tasks, metric choice plays a major role in our results, as widely used proxy scores can lead to different model rankings. We emphasize the importance of calibrating against ground truth to guard against metric hacking, such as over-warping in the spatial alignment example. Moreover, because execution traces record every exploratory check, failed attempt, and refinement, they enable systematic program distillation---compiling effective heuristics discovered by frontier models into fast, deterministic, and open-source scientific software.

Several limitations should be considered when interpreting our results. First, we focused on spatial transcriptomics alignment; extending our publication-to-benchmark pipeline to more problems, especially open-ended biological analyses, will require domain-specific data contracts and validation checks. Second, because frontier models and command-line harnesses evolve rapidly, point-in-time rankings reflect current default vendor configurations and might change as new models are released and older ones are deprecated. This is in fact a primary motivation for our exploration of program synthesis as a path toward reproducible, vendor-independent tools. Third, although more exhaustive hyperparameter tuning of baseline packages or prompt engineering could further improve individual scores, our evaluation intentionally reflects the practical scenario in which scientists apply these tools with a limited tuning budget (or, as in our case, with agent-assisted tuning).

To support transparent auditing and community benchmark development, we publicly release the Paper2Bench pipeline, all tasks, the containerized evaluation harness, and an interactive trace explorer containing all agent logs. More broadly, rather than viewing AI agents solely as black-box replacements for human analysts, we envision a collaborative paradigm in which auditable benchmarks and trace distillation harness agent exploration to build more rigorous, interpretable, and reusable tools.

%% file: sections/availability.tex
\section*{Data availability}
All data are available at Zenodo (\url{https://zenodo.org/records/XXXXXXX}), and traces can be explored at \url{https://ericzhy0815.github.io/agent-trace-explorer/}.

\section*{Code availability}
 Code to reproduce our analysis is available at \url{https://github.com/Y-Chen-LAB/spatial-alignment-agent}.

%% file: sections/acknowledgements.tex
\section*{Acknowledgements}
We acknowledge the Joint High Performance Computing Exchange (JHPCE) at the Johns Hopkins Bloomberg School of Public Health for computational support. This work was supported in part by a Gemini Academic Research Program and a Google Academic Research Award to Y.C. AI tools helped edit the text of this manuscript and write analysis and plotting code; the authors reviewed the final output and take full responsibility for the content and integrity of the work.

%% file: sections/supplementary.tex
\clearpage
\section*{Supplementary Information}
\setcounter{figure}{0}\setcounter{table}{0}
\renewcommand{\figurename}{Supplementary Figure}
\renewcommand{\tablename}{Supplementary Table}

\paragraph{Contents}
\begin{itemize}\setlength{\itemsep}{0pt}
  \item Supplementary Note 1: Task construction details (Supplementary Tables~\ref{tab:methods-families}, \ref{tab:tasks} and~\ref{tab:methods-albis})
  \item Supplementary Note 2: Published methods rerun under one scorer (Supplementary Figure~\ref{fig:supp-methods})
  \item Supplementary Note 3: Agents, harness and prompts (Supplementary Table~\ref{tab:methods-agents})
  \item Supplementary Note 4: Leakage audit (Supplementary Figure~\ref{fig:supp-audit} and Supplementary Table~\ref{tab:leakage})
  \item Supplementary Note 5: Scoring details
  \item Supplementary Note 6: Statistical details
  \item Supplementary Note 7: Trace processing and content (Supplementary Figures~\ref{fig:supp-keywords} and~\ref{fig:supp-trace-content})
  \item Supplementary Note 8: Run-to-run variation and its sources (Supplementary Figures~\ref{fig:supp-variation} and~\ref{fig:supp-anova})
\end{itemize}
\begin{tabular}{@{}ll@{}}
  Supplementary Figure~\ref{fig:supp-methods} & Evaluation of published alignment packages and geometric baselines  \\ & under a unified scorer \\
  Supplementary Figure~\ref{fig:supp-audit} & Container isolation layers and automated leakage audit results \\
  Supplementary Figure~\ref{fig:supp-keywords} & Algorithmic steps explored and retained across models and their association  \\ & with alignment scores \\
  Supplementary Figure~\ref{fig:supp-trace-content} & Quantitative analysis of execution trace volume, embeddings,  \\ & self-evaluations, and vocabulary \\
  Supplementary Figure~\ref{fig:supp-variation} & Run-to-run reproducibility across independent replicates \\
  Supplementary Figure~\ref{fig:supp-anova} & Variance decomposition of alignment accuracy, cost, and runtime \\
  Supplementary Table~\ref{tab:methods-families} & Task families and data sources \\
  Supplementary Table~\ref{tab:tasks} & Tissues, spatial technologies, and evaluation metrics per task family \\
  Supplementary Table~\ref{tab:methods-albis} & Simulation parameters for synthetic tissue tasks \\
  Supplementary Table~\ref{tab:methods-agents} & Agent harnesses, models, and launch configurations \\
  Supplementary Table~\ref{tab:leakage} & Modes of benchmark leakage observed prior to container isolation \\
\end{tabular}

\section*{Supplementary Note 1: Task construction details}
\paragraph{From publication to executable task.} Paper2Bench compiles a source publication into benchmark tasks across ten automated stages: source ingestion; extraction of experimental task units from figures, tables, methods, code, and datasets; normalization of inputs into typed data contracts; inference or specification of output schemas; attachment of deterministic scoring profiles; generation of blind task prompts; construction of isolated workspaces; automated validation; task prioritization by discriminative power, scientific relevance, and computational cost; and emission of runner metadata. Large language models assist in drafting candidate task extractions, output contracts, prompts, and rankings, while deterministic checks govern acceptance across six lifecycle states (\emph{draft}, \emph{executable}, \emph{validated}, \emph{calibrated}, \emph{sealed}, and \emph{published}).

Automated validators written by AI agents verify schema compliance, file checksums, data licensing and provenance, workspace reproducibility, submission format validation (including negative test fixtures), baseline execution (identity transform, geometric fit, and at least one published alignment method), metric determinism and perturbation sensitivity, absence of leaked metadata, and scorer integrity.

\paragraph{Dataset curation.} All tasks are generated programmatically from a declarative registry. Domain-specific curation was performed across task families to ensure exact fidelity to the source publications: matching author notebook scoring modes and marker gene lists (DLPFC); recompiling Bregma slice sets, filtering ambiguous cells, and harmonizing anatomical region labels (MHPR); reconstructing slice assignments (MOp); applying published cell quality filters (whole-brain MERFISH); standardizing input coordinate frames (\emph{Drosophila} embryo); regenerating expression matrices with per-section centering and updated marker lists (Open-ST); restoring raw count matrices and linking perturbation metadata (MBSA and MBSP); anonymizing cell identifiers and permuting row order (STARmap and BARISTA-seq); and constructing lattice-corrected variants for simulated spot tasks. 

\paragraph{Coordinate unit mismatch in the published \emph{Drosophila} embryo evaluation.} Rebuilding the 16-section \emph{Drosophila} E14--16 embryo Stereo-seq stack (\texttt{sa-019})~\citep{wang2022flysta3d} from raw SABench inputs uncovered a 20-fold scale discrepancy between the input coordinates and the ground-truth coordinates used in the published evaluation~\citep{yan2026sabench}. In SABench, input sections are centered raw Stereo-seq pixel coordinates (\texttt{raw\_x}, \texttt{raw\_y}; spatial span $\sim 600$ units), whereas the reference ground-truth positions (\texttt{new\_x}, \texttt{new\_y}) are stored in Flysta3D 3D-atlas units that are exactly 20-fold smaller ($\text{scale} = 0.05$, plus a small residual rigid rotation per section). Because SABench computed mean absolute error ($\mathrm{MAE}$) directly between unscaled output coordinates and $0.05\times$-scaled ground-truth coordinates (SABench Fig.~5f and Supplementary Table \texttt{Flysta3D\_MAE}), even the stationary reference slice (\texttt{s0}, which no alignment method transforms) incurred an identical floor error of $102.29$ across all ten evaluated methods (reproduced to six decimal places by our scorer prior to rescaling). 

Consequently, under the unscaled evaluation, nearly all rigid and optimal-transport baselines and agent runs compressed tightly around the unaligned identity error, whereas methods that shrank coordinates toward the origin appeared artificially accurate (\texttt{moscot\_warp}, $39.8$; \texttt{cast\_rescale\_affine}, $88.5$). Rescaling the ground-truth coordinates by $20\times$ into native input pixel units resolved this artifact: the stationary reference slice has an $\mathrm{MAE}$ of $0$, unaligned inputs have an $\mathrm{MAE}$ of $166.3$, and alignment methods separate cleanly—led by PASTE ($42.3$, a 75\% error reduction), moscot ($48.0$, 71\%), \texttt{claude-sonnet-5.5} ($55.4$, 67\%), SPACEL ($78.4$, 53\%), \texttt{claude-opus-5.5} ($118.7$, 29\%), and \texttt{gpt-6-astra} ($122.4$, 26\%), while shrinkage configurations (\texttt{cast\_rescale\_affine}, $217.3$) and over-warped nonrigid runs score worse than unaligned inputs. 

\paragraph{Sources and task families.} The five task families and their data sources are summarized in Supplementary Table~\ref{tab:methods-families}, with tissue types, spatial platforms, and available evaluation metrics detailed in Supplementary Table~\ref{tab:tasks}. Of the 240 tissue sections in SABench, 236 were reconstructed exactly or near-exactly (100 exact, 136 near-exact, 2 with modified quality filters, and 2 from alternative public releases), covering all 295 evaluated section pairs (260 experimental pairs and 35 synthetic rotation and cropping perturbations). 

Four whole-brain MERFISH tasks use raw imaging coordinates registered to the Allen Common Coordinate Framework (CCF) by the original authors; each section shows an isotropic scaling factor of $1.17 \pm 0.05$ and about 250~\textmu m of local nonlinear deformation, making them the most challenging geometric reconstructions in the benchmark. Ground-truth coordinates derive from simulation (12 tasks), known synthetic transformations of experimental sections (11), author-curated registrations (5), and anatomical atlas coordinates (4). All underlying datasets are publicly available: SABench data (Zenodo \texttt{10.5281/zenodo.18605715}), GEASO data (Zenodo \texttt{18811760}), Open-ST (GEO \texttt{GSE251926})~\citep{schott2024openst}, spatial CUT\&Tag-RNA (GEO \texttt{GSE205055})~\citep{zhang2023cuttag}, RIBOmap (Single Cell Portal \texttt{SCP1835})~\citep{zeng2023ribomap}, STARmap PLUS (\texttt{SCP1375})~\citep{zeng2023starmapad}, and the Albis simulator~\citep{albis} (MIT license).

Ground-truth coordinates are available for the three-dimensional SABench stacks, all simulated tasks, and most GEASO tasks (32 tasks total); cell-type or regional labels are available for annotated SABench tasks and two cross-modality pairs (69 tasks); and all remaining tasks are evaluated via gene expression agreement.

\begin{table}[!htbp]
  \centering
  \caption{\small\textbf{Task families and sources.} Abbreviations: DLPFC, dorsolateral prefrontal cortex; BCA, breast cancer; MBSA and MBSP, mouse brain sagittal anterior and posterior; AMBC, adult mouse brain coronal; MHPR, mouse hypothalamic preoptic region; SCC, squamous cell carcinoma; MOp, primary motor cortex; MOB, mouse olfactory bulb; WMB, whole mouse brain.}
  \label{tab:methods-families}
  \small
  \begin{tabularx}{\linewidth}{@{}p{0.2\linewidth}rp{0.16\linewidth}X@{}}
    \toprule
    Family & Tasks & Source & Content \\
    \midrule
    Serial sections (sa-0xx) & 35 & SABench~\citep{yan2026sabench} & DLPFC Visium~\citep{maynard2021dlpfc}, BCA, MBSA and MBSP, AMBC Xenium~\citep{janesick2023xenium}, MHPR MERFISH~\citep{moffitt2018mhpr}, SCC~\citep{ji2020scc}, BARISTA-seq~\citep{sun2021baristaseq}, STARmap~\citep{wang2018starmap}, fly embryo (16 sections)~\citep{wang2022flysta3d}, MOp MERFISH (31 and 33 sections)~\citep{zhang2021mop}, Open-ST lymph node (19 sections)~\citep{schott2024openst}, WMB MERFISH (129 sections and windows of 5 and 10)~\citep{zhang2023wmb}, MOB Stereo-seq~\citep{chen2022stereoseq} and Slide-seqV2~\citep{stickels2021slideseqv2}, and four rotation and crop sweeps \\
    Motor cortex series (sa-1xx) & 12 & SABench~\citep{yan2026sabench} & MOp MERFISH, one task per sample of two mice, 4 to 6 sections each \\
    Cross-platform pairs (sa-2xx) & 45 & SABench~\citep{yan2026sabench} & All pairs of ten platforms on adult mouse brain sections: ST~\citep{stahl2016st}, Visium, Visium HD~\citep{oliveira2025visiumhd}, Xenium, Xenium 5K, MERFISH~\citep{chen2015merfish}, STARmap PLUS~\citep{shi2023starmapplus}, Slide-seq, Stereo-seq and CosMx~\citep{he2022cosmx} \\
    Simulated tissue (sa-3xx) & 12 & Albis simulator~\citep{albis} & Simulated serial sections with known positions (Supplementary Table~\ref{tab:methods-albis}) \\
    Rotation, crop, 3D (sa-4xx) & 17 & GEASO data~\citep{wang2026geaso}, WMB MERFISH & STARmap PLUS depth, overlap and stitching~\citep{shi2023starmapplus}; STARmap and RIBOmap cross-modality~\citep{zeng2023ribomap}; cross-animal and rotation sweep~\citep{zeng2023starmapad}; spatial CUT\&Tag-RNA~\citep{zhang2023cuttag}; 129-section MERFISH hemibrain; Open-ST reconstruction; whole-brain MERFISH from raw coordinates (129, 33, 17 and 33 irregular sections) \\
    \bottomrule
  \end{tabularx}
\end{table}

\begin{table}[!htbp]
  \centering
  \caption{\small\textbf{Tissues, technologies and available scores per family.} For each family, the tissues and technologies it covers and how many tasks can be scored with true positions, with labels but no true positions, or with expression agreement only. The serial-sections family also holds some cross-platform pairs and three-dimensional stacks from SABench.}
  \label{tab:tasks}
  \scriptsize
  \IfFileExists{tables/task_families.tex}{\input{tables/task_families.tex}}{\fbox{Table in preparation}}
\end{table}

\paragraph{Simulated tissues.} The 12 synthetic tasks were generated using the Albis simulator across 6 spatial domains, 8 cell types, and 10 serial slabs per tissue (Supplementary Table~\ref{tab:methods-albis}). Each section undergoes a random rigid transformation with rotations up to $\pm 270^\circ$ and translations up to half the tissue radius. Section-specific batch effects apply a gene-wise log-fold perturbation drawn with standard deviation $\sigma$, followed by Poisson count resampling. Ground-truth positions are defined in the coordinate frame of the first section, with observation rows randomly permuted and cell identifiers anonymized.

Validation on these synthetic datasets revealed two notable failure modes of proxy alignment metrics. First, grid-based expression correlations exhibit a near-zero upper bound under high Poisson sparsity (split-half correlation of 0.037 on a $10 \times 10$ spatial grid). Second, nearest-neighbor label metrics can favor imprecise alignments: a coarse centroid fit with a 371~\textmu m coordinate error achieved a label matching accuracy of 0.837, exceeding the score of the exact ground-truth registration (0.815). Coordinate error is therefore used as the primary evaluation metric across all simulated tasks.

\begin{table}[!htbp]
  \centering
  \caption{\small\textbf{Simulated tasks.} $R$, tissue radius; $\sigma$, standard deviation of the batch effect.}
  \label{tab:methods-albis}
  \small
  \begin{tabularx}{\linewidth}{@{}p{0.17\linewidth}p{0.33\linewidth}X@{}}
    \toprule
    Task & Regime & Key parameters \\
    \midrule
    sa-301 & strong mixing, cells & 24,207 cells, $R=2{,}050$~\textmu m, $\sigma=1.5$ \\
    sa-302 & Xenium-like cells & 600,000 cells, $R=6{,}000$~\textmu m, $\sigma=1.5$ \\
    sa-303 & Visium spots & 55~\textmu m spots on a 100~\textmu m lattice, $\sigma=0.3$ \\
    sa-304, sa-305 & Visium HD bins & 16~\textmu m ($\sigma=0.7$) and 8~\textmu m ($\sigma=0.8$) \\
    sa-306, sa-307 & as sa-303 and sa-304, lattice fixed to the slide & removes the lattice leak \\
    sa-308 & cells with gaps & 30 slabs of 137~\textmu m, 9 kept at irregular spacing \\
    sa-309, sa-310 & sparse Visium, without and with gaps & 120,000 cells; spots with no counts or fewer than 3 genes dropped \\
    sa-311 & Visium HD 32~\textmu m, low signal & $\sigma=1.2$, weak landmarks \\
    sa-312 & decoy flip & mirror-image decoy domains; half of the non-reference sections mirrored \\
    \bottomrule
  \end{tabularx}
\end{table}

\paragraph{Task package structure.} Each task package consists of a machine-readable task card and natural-language task instructions. The task card specifies the ordered input sections, section-pair evaluation topology (sequential adjacent pairs or star-graph comparisons against a reference section), withheld annotation columns, anatomical regions for alpha-shape overlap, marker gene sets (published marker lists of 117 genes for DLPFC and 155 for MHPR, or up to 100 top spatially variable markers otherwise), and the scoring profile. 

Agents are instructed to output one CSV file per section containing columns \texttt{obs\_id}, \texttt{x}, and \texttt{y} and submissions are validated to ensure exact observation ID matching, row ordering, and finite numerical values. 
\section*{Supplementary Note 2: Published methods rerun under one scorer}
\begin{figure}[!htbp]
  \centering
  \figorplaceholder{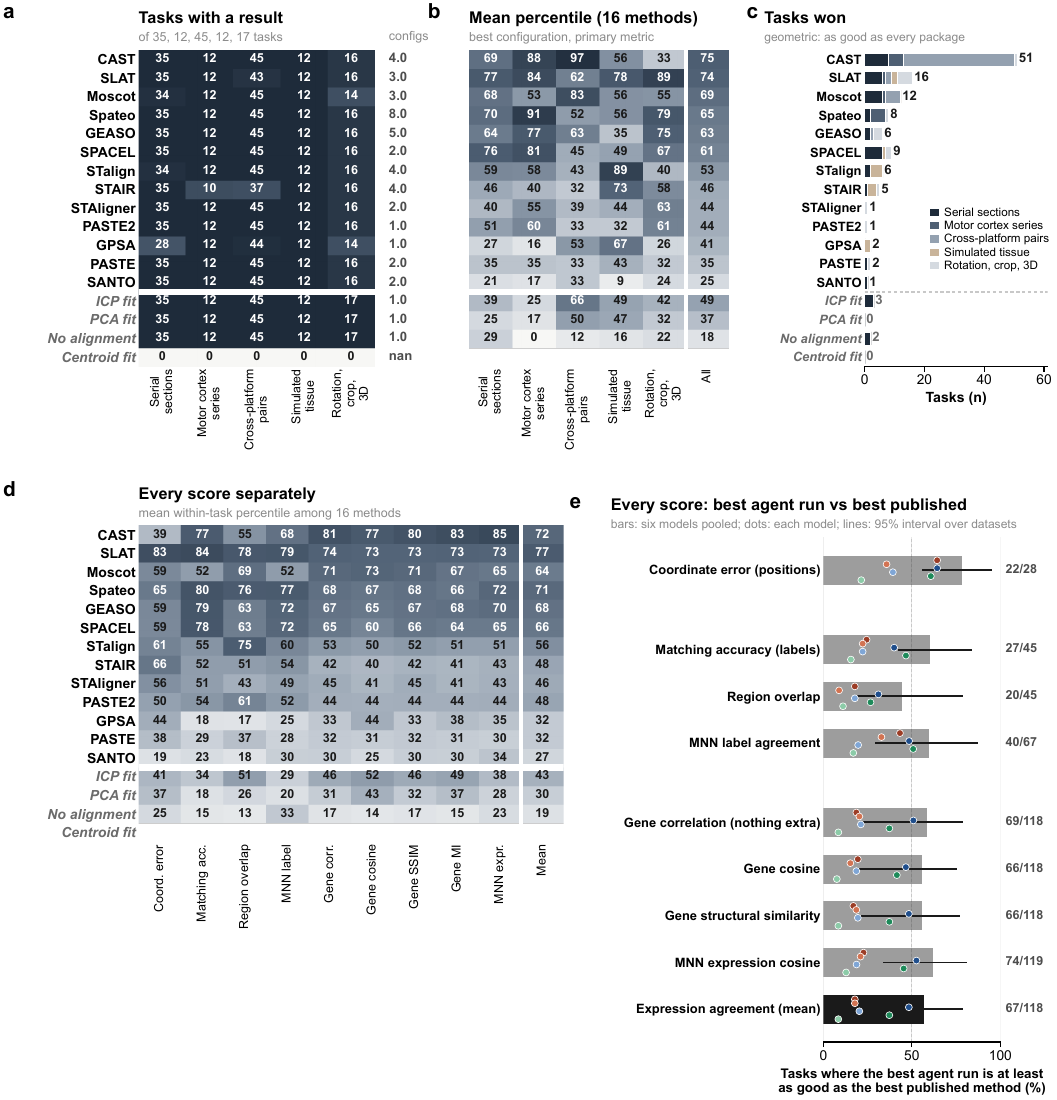}{\linewidth}
  \caption{\small\textbf{Evaluation of published alignment packages and geometric baselines under a unified scorer.}
  \textbf{a}, Task coverage across families and number of parameter configurations evaluated per method.
  \textbf{b}, Mean within-task percentile across the 16 reference baselines on the primary metric, stratified by task family.
  \textbf{c}, Number of tasks on which each published package achieved the highest baseline score, or on which a geometric baseline matched or outperformed all packages.
  \textbf{d}, Mean within-task percentile stratified by individual evaluation metric. MNN, mutual nearest neighbors; SSIM, structural similarity; MI, mutual information.
  \textbf{e}, Percentage of tasks on which the best basic-prompt agent run matched or outperformed the best published configuration across individual metrics. PCA, principal component analysis; ICP, iterative closest point.}
  \label{fig:supp-methods}
\end{figure}

With Claude Code (Opus 5.5 with high effort), we re-executed 13 published alignment packages across 41 parameter configurations alongside three geometric baselines on the exact task inputs and scoring harness used for agent evaluation, taking the top-performing configuration of each package on a given task as its representative score (Supplementary Figure~\ref{fig:supp-methods}). Package versions were \texttt{paste-bio} 1.4.0 (\texttt{POT} 0.9.7), \texttt{paste2} 1.0.1, \texttt{spateo-release} 1.1.1, \texttt{SPACEL} 1.1.8, \texttt{GEASO} (commit \texttt{2381513}), \texttt{CAST} 0.4 (commit \texttt{3ebcf4f}), \texttt{STAligner} 1.0.0 (commit \texttt{85ca4ce}), \texttt{scSLAT} 0.3.0, \texttt{STalign} 1.0 (commit \texttt{b2068ed}), \texttt{STAIR-tools} 1.3.1, \texttt{SANTO} 0.0.3, \texttt{GPSA} 0.6.15, and \texttt{moscot} 0.5.2 (\texttt{ott-jax} 0.6.0). Evaluated configurations varied transformation families (rigid versus nonrigid), reflection search, coordinate rescaling, initialization schemes, and feature selection; 26 core configurations were executed across at least 104 tasks. Two SPACEL configurations that required ground-truth region annotations and a fourth geometric baseline that fit transformations to ground-truth cell-type centroids were excluded from baseline comparisons. Coverage was nearly complete across families (Supplementary Figure~\ref{fig:supp-methods}a).

Across all 16 reference methods, CAST achieved the highest mean within-task percentile (75th) and SLAT ranked second (74th), with leading methods varying by task family: SLAT led on serial sections and on rotation, cropping, and 3D reconstruction tasks, whereas STalign led on simulated tissues (Supplementary Figure~\ref{fig:supp-methods}b). Simple geometric baselines proved competitive against several specialized tools: ICP registration achieved a mean within-task percentile of 49 (outperforming six published packages) and PCA similarity alignment achieved 37 (outperforming two).

CAST ranked as the single best published method on 51 of 120 tasks (including 37 cross-platform pairs), while geometric baselines matched or exceeded all published packages on 4 tasks (Supplementary Figure~\ref{fig:supp-methods}c). Stratified by individual metrics, SLAT led on all coordinate- and label-based scores, whereas CAST led on all gene expression similarity scores (Supplementary Figure~\ref{fig:supp-methods}d). Hyperparameter configuration substantially affected baseline performance: the best and worst configurations of SLAT, moscot, and CAST differed by a median of 24 to 48 percentile points within a task. Across individual evaluation metrics, the best of the six basic-prompt agent runs matched or outperformed the top published configuration on 44\% (region overlap) to 62\% of tasks (Supplementary Figure~\ref{fig:supp-methods}e).

\paragraph{Comparison with originally reported benchmark values.} We compared our re-evaluated baseline scores against published source tables across 312 method-by-task entries from GEASO (1,567 source records covering rotation sweeps, stitching tasks, 129-section MERFISH reconstruction, and 19-section Open-ST reconstruction, converted at 0.0946~\textmu m per pixel) and SABench (nine tasks with method-level tables). Defining numerical reproduction as agreement within 0.02 (close: 0.06) for bounded metrics, $5^\circ$ ($20^\circ$) for rotation angles, and 15\% (40\%) for coordinate error, 63 of 312 comparisons were reproduced closely, 51 were within the broader tolerance, 127 differed, and 71 were not directly comparable. Differences on GEASO tasks primarily reflect GEASO's post-hoc Procrustes alignment prior to computing Euclidean pixel error, whereas our scorer evaluates direct per-axis coordinate error in native spatial units without post-hoc registration; where directly comparable, competing baseline packages performed at least as well in our reruns as reported in the original study.

\section*{Supplementary Note 3: Agents, harness and prompts}
\paragraph{Agent configurations.} Agent CLI versions, model identifiers, reasoning effort parameters, and launch commands are listed in Supplementary Table~\ref{tab:methods-agents}. To support \texttt{claude-opus-5.5} and \texttt{claude-sonnet-5.5}, an updated Claude Code binary was mounted read-only into the container image. OpenAI models were accessed via Codex CLI.

\begin{table}[!htbp]
  \centering
  \caption{\small\textbf{Agent harnesses, models, and launch configurations.} Software versions correspond to the container environment; launch commands are abridged.}
  \label{tab:methods-agents}
  \scriptsize
  \begin{tabularx}{\linewidth}{@{}p{0.15\linewidth}p{0.18\linewidth}p{0.07\linewidth}X@{}}
    \toprule
    Agent software & Model & Effort & Launch (abridged) \\
    \midrule
    Claude Code 2.1.281 & \texttt{claude-opus-5.5} & high & \texttt{claude -{}-print -{}-output-format stream-json -{}-model <m> -{}-dangerously-skip-permissions -p "<prompt>"} \\
     & \texttt{claude-sonnet-5.5} & default & as above \\
    Codex CLI 0.154.0 & \texttt{gpt-6-astra} & xhigh & \texttt{codex exec -{}-json -{}-model <m> -c model\_reasoning\_effort=xhigh -{}-dangerously-bypass-approvals-and-sandbox "<prompt>"} \\
     & \texttt{gpt-5.5} & xhigh & as above \\
    Gemini CLI 0.60.0 & \texttt{gemini-4-argon} & none & \texttt{gemini -{}-prompt "<prompt>" -{}-output-format stream-json -{}-model <m> -{}-yolo} \\
     & \texttt{gemini-3.5-flash} & none & as above \\
    \bottomrule
  \end{tabularx}
\end{table}

\paragraph{Prompting conditions.} For representative task \texttt{sa-001}, the basic prompt comprises 4,056 characters and the rich-context prompt 21,170 characters (full texts provided below). Under the rich-context condition, isolated virtual environments were pre-configured for 11 packages: PASTE, PASTE2, Spateo, SPACEL, STAIR, GPSA, STalign, SLAT, CAST, STAligner, and SANTO (excluding GEASO and moscot). The rich-context prompt also included a modality-specific recommendation table adapted from SABench (for example, prioritizing SPACEL and PASTE2 for Visium serial sections). Across the six models evaluated under both prompts, the change in within-task percentile for \texttt{gemini-3.5-flash} yielded $P = 0.22$ after Holm adjustment, while gains for \texttt{claude-opus-5.5} were concentrated in a subset of datasets. On tasks with ground-truth positions, the geometric mean ratio of coordinate error (rich context relative to basic prompt) was 0.64 for \texttt{claude-opus-5.5} and 0.74 for \texttt{claude-sonnet-5.5}, compared with 1.59 for \texttt{gpt-6-astra}, 1.79 for \texttt{gpt-5.5}, 1.89 for \texttt{gemini-3.5-flash}, and 3.86 for \texttt{gemini-4-argon} (95\% CI: 0.54--130). Rich context also increased tool-call volume (for example, from a median of 17 to 42 calls for \texttt{claude-opus-5.5}) and reduced the use of custom numerical optimization in \texttt{gpt-6-astra} by 28 percentage points.

\paragraph{Container harness.} All evaluations were executed under Singularity 3.11.4 (\texttt{-{}-containall -{}-cleanenv -{}-no-home}) using a base image built on \texttt{python:3.11-slim} with Node.js 22, \texttt{uv}, and the three vendor CLI tools. Only the task workspace and an ephemeral state directory were writable; software mounts and API credentials were read-only, and environment variables containing keys were scrubbed prior to agent startup. Standard tasks were allocated 4 CPU cores, 48~GB RAM, a 360-min agent execution limit, and a 60-min scoring window (7~h total wall time); large 3D reconstruction tasks were allocated 4 CPU cores, 96~GB RAM, a 720-min agent limit, and a 360-min scoring window (20~h total wall time), with no GPU access. Across all scored runs, 95\% completed within 71~min (129~min for large 3D tasks), with a maximum runtime of 361~min. Upon termination, workspace outputs and execution logs were archived for auditing and the container workspace was purged.

\paragraph{Prompts.} Verbatim prompt templates recorded by the execution harness are reproduced below: the basic prompt (illustrated for task \texttt{sa-001}), the rich-context package documentation and modality recommendation tables, and the strategy-distillation prompt provided to the synthesizer agent.

\paragraph{Basic prompt} (task sa-001, complete; line breaks added).
{\scriptsize
\begin{verbatim}
You are running inside a blind benchmark workspace.

Read TASK.md, inspect only the data files under data/, and write the required coordinate CSVs under
submission/aligned_coordinates/.
No local scorer is available during solving. Do not search parent directories, repository paths, prior
runs/results, benchmark source files, scorer code, Downloads caches, or any location outside this workspace.
Allowed inputs are the sanitized H5AD expression data, observation IDs, `obsm/spatial` coordinates, and
slice identities described in TASK.md. Do not use or search for hidden aligned coordinates, ground-truth
transforms, perturbation metadata, scorer labels, evaluation landmarks, split manifests, or private source
paths.
You may create scripts or notes inside this workspace, but do not modify source data.
When finished, give a concise final answer describing the submitted files.

TASK.md:
# Agent Task: sa-001-dlpfc-serial-align

# DLPFC Serial Alignment

Align the four 10x Visium DLPFC slices into a common coordinate system.

Submit one coordinate CSV per slice with columns `obs_id,x,y`. The row order and `obs_id` values must match
the source `.h5ad` file.

Note: source-manifest details that referenced scorer-only annotations, ground-truth coordinates, or
transform metadata were omitted from this agent prompt.

## Data Visibility

The H5AD files under `data/` are sanitized agent-visible inputs.

Allowed inputs:

- Observation IDs from `obs/_index`
- Gene expression from `X`, `var`, and any provided expression `layers`
- Original two-dimensional coordinates from `obsm/spatial`
- Slice identities and required output filenames from this task prompt

Not allowed and not present in the solve workspace:

- Hidden gold aligned coordinates or 3D atlas coordinates
- Ground-truth transforms, perturbation metadata, or perturbation configuration files
- Evaluation landmarks, scorer-only split manifests, or local scorer source
- Anatomical or scoring labels reserved for evaluation
- Private benchmark source paths, local download caches, previous run artifacts, or benchmark repository
  files

Use only the allowed inputs above. Do not infer, search for, or rely on hidden source paths, scorer
internals, prior submissions, or withheld metadata.

## Data

The input `.h5ad` files are available in this workspace:

- `151673`: `data/151673.h5ad`
- `151674`: `data/151674.h5ad`
- `151675`: `data/151675.h5ad`
- `151676`: `data/151676.h5ad`

## Required Output

Create coordinate CSV files in:

- `submission/aligned_coordinates/151673.csv`
- `submission/aligned_coordinates/151674.csv`
- `submission/aligned_coordinates/151675.csv`
- `submission/aligned_coordinates/151676.csv`

Each CSV must contain exactly these columns:

```text
obs_id,x,y
```

Rules:

- `obs_id` values and row order must match the source `.h5ad` observations.
- `x` and `y` must be finite numeric aligned coordinates.
- Do not modify the source data files.
- You may create code, logs, figures, or a short `submission/report.json`, but scoring uses the coordinate
  CSV files.

## Execution Strategy

First create a simple valid submission for every required output CSV, even if it only uses a conservative
baseline transform. Then improve methodically: try richer custom models or refinements only after a complete
returnable answer is already on disk. Whenever you find a better alignment, overwrite the same prediction
CSVs. Keep the fallback CSVs valid while you explore stronger approaches.

## Local Scoring

No local scorer is available in this solve workspace. Final scoring is run by the benchmark harness after
your agent process exits.
Do not search parent directories, repository paths, previous runs, benchmark source files, scorer code,
Downloads caches, or other locations outside this workspace.

For Python, use the workspace helper:

```bash
./agent_python.sh your_script.py
# or
./agent_python.sh -c "import h5py; print('ok')"
```

The helper runs `uv` with common cached analysis packages: `numpy`, `scipy`, `pandas`, `scikit-learn`, and
`h5py`.
\end{verbatim}
}

\paragraph{Method priorities added to the rich-context prompt} (verbatim; \nolinkurl{references/prompt_assets/spatial_alignment_methods_paper_prior.md} in the code repository, last changed 2026-09-22). The rich-context prompt also lists the package environments with an API example for each and the documentation menu.
{\scriptsize
\begin{verbatim}
## Benchmark Context (Paper-Informed Package Priority)

The benchmark paper compared multiple spatial-alignment method families on
related spatial-transcriptomics tasks. In this prompt variant, you are given
high-level paper-derived method priorities so you can spend your package-testing
budget on methods that were strongest in the paper. These are method priors,
not task labels, not ground-truth transforms, and not local scorer information.

You still must use only the allowed task inputs: sanitized AnnData/CSV files,
spot IDs, gene expression, original `obsm/spatial` x/y coordinates, and slice
identities from `TASK.md`. Do not search for or use paper result tables, hidden
aligned coordinates, benchmark source, scorer code, prior submissions, or
withheld labels.

First create a complete valid submission for every required output CSV. Then
try package-backed methods in a paper-informed order. A useful package attempt
means using the method-specific Python executable from `.method-cache/METHOD_ENVS.md`,
running an alignment API on the task data or a representative pair/subset, and
recording whether it produced usable coordinates. If a high-priority method
fails to import, has incompatible API/runtime behavior, exceeds a deliberate
timeout, or looks worse under allowed input-only diagnostics, move on and keep
the best current submission. Do not treat a quiet import as an immediate
failure: several Torch/numba-heavy method envs can spend a minute or two in
first-import/JIT startup. Wait unless the process errors or your bounded timeout
is genuinely too expensive for the task.

### Paper-Informed Priority By Task Family

| Task family / data shape | Try first | Then try | Notes |
|---|---|---|---|
| DLPFC-like Visium serial alignment | `spacel`, `paste2` | `cast`, `stair`, `spateo` | Direct paper DLPFC
  gene+landmark tables favored `SPACEL` and `PASTE2`; `CAST` and `STAIR` were also strong. |
| MHPR/MERFISH or image-like serial alignment | `spateo`, `spacel` | `paste2`, `stair`, `slat` | Direct paper MHPR
  tables favored `spateo` rigid/non-rigid and `SPACEL`; `PASTE2`, `STAIR`, and `SLAT` were competitive. |
| Generic NGS/Visium/STereo-seq slices | `paste2`, `spacel`, `spateo` | `stair`, `paste` | Aggregate NGS rankings
  favored `PASTE2`; `SPACEL` and `spateo` were consistently strong. `PASTE` is a useful faster fallback. |
| 3D serial / MAE-style stacks | `spacel` | `paste`, `paste2`, `stair`, `spateo` | Paper Flysta3D MAE favored
  `SPACEL`, then PASTE-family / `STAIR` / `spateo` methods. |
| MLN-like multi-section MAE tasks | `spateo` rigid | `slat`, `spateo` non-rigid, `stalign`, `spacel` | Paper MLN MAE
  favored `spateo` rigid and `SLAT`. |
| Unknown or mixed platform | `spateo`, `spacel`, `paste2` | `stair`, `cast`, `slat`, `paste` | Start with the robust
  paper winners, then adapt based on data size and import/API success. |

Do not treat this table as a hard rule. It is a search prior. If the dataset has
very large slices, use subsets to estimate transforms first. If a method needs
domain labels that are not present, you may create unsupervised domains from
allowed expression/coordinate data, but do not invent or recover withheld
anatomical labels.

### Practical Package Guidance

- Prefer `spateo` and `SPACEL` when tissue geometry, non-rigid deformation, or
  image-like spatial continuity looks important.
- Prefer `PASTE2` when partial overlap, NGS/Visium-style slices, or robust
  optimal-transport matching looks important.
- Keep `PASTE` as a simpler/faster fallback, especially for large serial stacks
  where full package methods are too slow.
- Use `CAST`, `STAIR`, or `SLAT` when their APIs are importable and the task
  looks compatible with embedding or graph matching.
- Treat `GPSA`, `STAligner`, `STalign`, and `SANTO` as optional method-specific
  attempts rather than first defaults unless their API/docs clearly fit the
  task.
- For slower method families, run a small representative pair/subset first and
  allow enough time for first import/JIT startup before deciding the method is
  infeasible.

### Required Reporting

In `submission/report.json` or a local note, record:

- which paper-prior methods you attempted,
- which package env executable you used,
- whether the package produced usable transformed coordinates,
- why you moved on from any failed high-priority method,
- which candidate you finally submitted.
\end{verbatim}
}

\paragraph{Strategy distillation.} We partitioned 119 tasks (excluding \texttt{sa-005} and \texttt{sa-023}) into five folds using a fixed random seed under both a random split (23 or 24 tasks per fold) and a task-family split. For each fold, an isolated training directory contained the final scripts and scores of basic-prompt runs from \texttt{claude-opus-5.5} and \texttt{gpt-6-astra} on the four training folds, the sanitized training inputs, and the geometric baseline implementations. 

The synthesis agent (Claude Code with \texttt{claude-opus-5.5}) executed inside a container with network access and produced a standalone alignment program, a methodological playbook, and synthesis notes (an ablation variant synthesized from training code without training scores was also generated). Each synthesized program was then evaluated on a clean blind workspace for every task in the held-out fold without LLM access.

\paragraph{Synthesizer prompt} (verbatim, cell random-train4-f0-scored; the other cells differ only in the numbers of training tasks and runs, and the code-only cells have no \texttt{scores/} folder).
{\scriptsize
\begin{verbatim}
# Synthesize a general alignment strategy

## The job

Earlier AI agents (claude-opus-5-5 and gpt-6-astra) each solved spatial-transcriptomics slice-alignment tasks
on their own. You have their records for 95 tasks (315 runs). Your job is to **synthesize what
they did into one general strategy** for aligning slices, and to package it so it can be applied to **new
tasks you have not seen**. Those new tasks may come from other tissues, platforms and task designs.

Do not try to beat any one run on any one task. The goal is the strategy that generalizes best: which
approaches the runs converged on, when each one is the right choice, how to recognize that from the inputs,
and how to combine or choose among them (an ensemble or decision rule).

## What you have (and nothing else)

- `runs/<task>/<model>__<job>/`: one directory per prior run: `solution/` (the code that produced
  its final submission, with `SOLUTION.md`) and `meta.json` (model, task, entry point). The runs' step-by-step
  traces are not included.
- `tasks/<task>/`: each training task's inputs exactly as the agents saw them (`TASK.md`, `data/*.h5ad`,
  `agent_python.sh`). Read-only: to try something on a task, copy the task directory into `scratch/` first.
- `reference_aligners/`: four simple aligners (identity, centroid similarity, PCA similarity, ICP
  similarity). `python3 reference_aligners/run_reference.py --help` shows how to run them on a workspace.
- `scores/` holds the benchmark score of every run and of every reference aligner on its training task (definitions in
  `scores/SCORES.md`). Use them to judge which strategies worked where.

There is no ground truth and no scorer here. The network reaches only the model API and the Python package
index (PyPI); nothing else is reachable, so do not try to look things up online.

## What to deliver, under `output/`

1. `output/policy/run.sh`: the strategy as a program. It is run as `bash output/policy/run.sh <workspace>`,
   where `<workspace>` has the same layout as `tasks/<task>/` (a `TASK.md`, `data/*.h5ad`, `agent_python.sh`
   and an empty `submission/aligned_coordinates/`). It must write the coordinate CSVs that `TASK.md` asks
   for into `<workspace>/submission/aligned_coordinates/`.
   - Decide everything from `TASK.md` and the data. Never look up a task by its name or ID, and never
     hard-code a per-task answer: the tasks it will meet are not in your training set.
   - Keep the policy self-contained under `output/policy/`: copy in whatever code you reuse from the runs or
     from `reference_aligners/`.
   - Python: `<workspace>/agent_python.sh` gives numpy, scipy, pandas, scikit-learn and h5py. If you need other
     public packages, list them in `output/policy/requirements.txt`: before `run.sh` starts they are installed
     from PyPI into the environment that `<workspace>/agent_python.sh` runs. The policy directory itself is
     read-only when it runs; write intermediates under `<workspace>/` (for example `<workspace>/work/`).
   - Budget per task: 8 CPU cores, 64 GB of memory, and the wall time in seconds in the environment variable
     `POLICY_TIME_BUDGET_S` (3 hours for most tasks). Plan for the larger inputs: some tasks have more than
     100 sections or more than a million cells in total.
   - Always leave a complete, valid submission: if a step fails or runs out of time, fall back to a cheaper
     method rather than leaving files missing.
2. `output/PLAYBOOK.md`: the same strategy in prose, written for another agent who will solve a new task by
   hand: what the runs learned, the decision rules (which approach when, and how to tell from the inputs),
   checks that catch a bad alignment without ground truth, and the mistakes to avoid.
3. `output/NOTES.md`: how you got there: which runs and tasks support each rule, what you tried, and how you
   tested the policy (it should run end to end on several training tasks of different kinds).

Work in `scratch/` for experiments, and keep it under 20 GB (delete copied task data you no longer need).

How to test, so that the work stays quick:
- Test end to end on small and medium training tasks only, each run capped at 10 minutes of wall time
  (`timeout 600 bash output/policy/run.sh <workspace>`). Do not run the policy in full on the largest tasks
  (many sections or over a million cells): check that it scales by timing it on a subset of their sections
  and extrapolating, and rely on the fallback for the rest.
- Run each test as one blocking command in the foreground and read its result when it returns. Do not start
  it in the background and poll it with repeated `sleep` and check commands: every check is another full turn.

When you finish, give a short summary of the strategy.
\end{verbatim}
}

\section*{Supplementary Note 4: Leakage audit}
\begin{figure}[!htbp]
  \centering
  \figorplaceholder{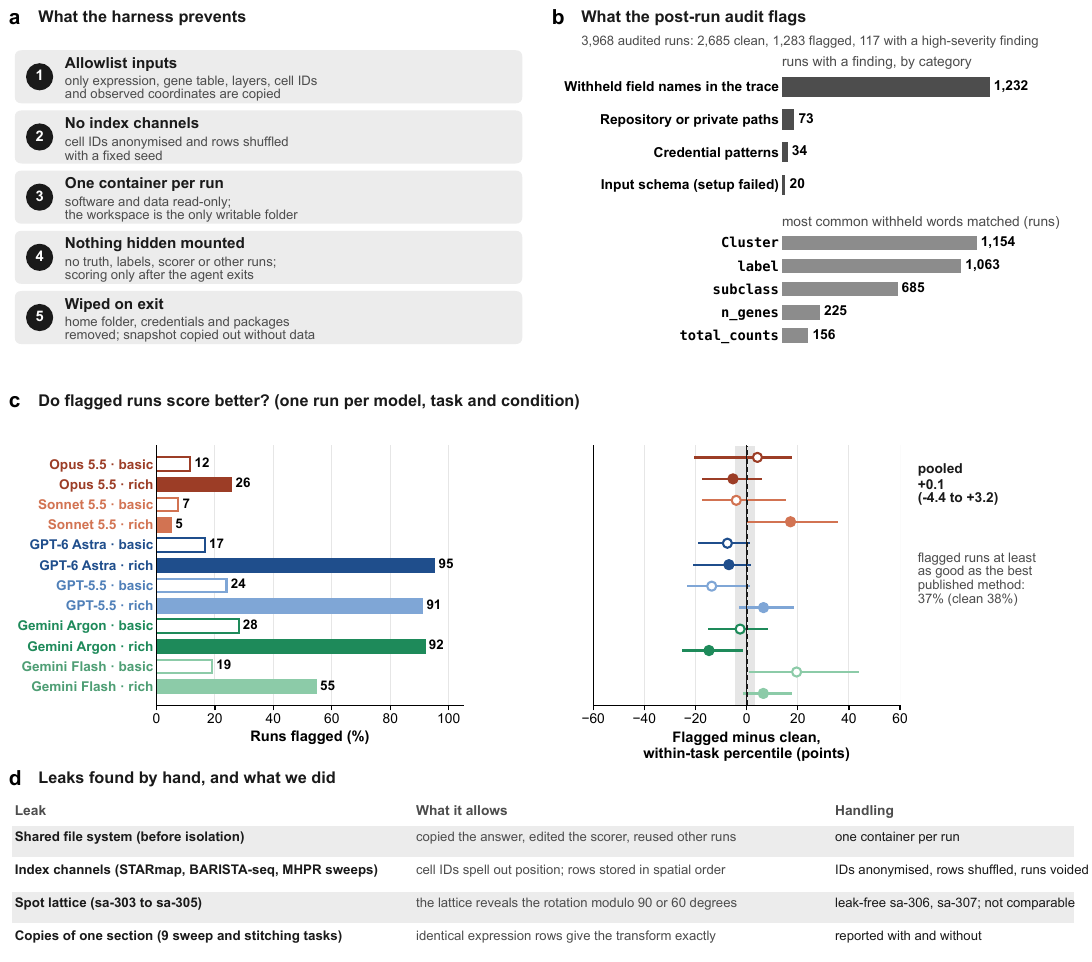}{\linewidth}
  \caption{\small\textbf{Container isolation layers and automated leakage audit results.}
  \textbf{a}, Isolation architecture preventing access to held-out data and scoring scripts inside the container workspace.
  \textbf{b}, Distribution of automated audit alerts across all audited runs by category and most frequently matched keywords.
  \textbf{c}, Proportion of flagged runs in the primary analysis set and within-condition percentile difference between flagged and unflagged runs.
  \textbf{d}, Summary of dataset-level shortcuts identified through trace inspection and remediation.}
  \label{fig:supp-audit}
\end{figure}

\paragraph{Leakage prevention.} We processed the input data to prevent held-out evaluation data to prevent entering the agent workspace (Supplementary Figure~\ref{fig:supp-audit}a). Workspace inputs are generated via a strict allowlist: across all 1,128 tissue sections in the benchmark, sanitized \texttt{h5ad} files retain only the expression count matrix, feature table, expression layers, anonymized observation identifiers, and observed spatial coordinates. 

\paragraph{Automated post-run auditing.} Following every run, we used a programmatic keyword-based audit to look through the execution trace, workspace directory, and submitted files. 

Across all runs in the primary analysis set ($n = 1{,}452$), 568 runs triggered keyword flags (generic biological terms such as ``Cluster'', ``label'' and ``subclass''), predominantly under the rich-context prompt, where pre-installed package documentation references standard AnnData annotation fields. Controlling for model and prompt condition, flagged runs showed no advantage over unflagged runs (mean difference of $+0.1$ within-task percentile points, 95\% CI: $-4.4$ to $+3.2$; Supplementary Figure~\ref{fig:supp-audit}), providing no evidence of answer leakage.

\paragraph{Dataset-level shortcuts identified via trace review.} Manual inspection of execution traces and task inputs uncovered several dataset-intrinsic shortcuts and guided benchmark remediation (Supplementary Figure~\ref{fig:supp-audit}d). 

On nine synthetic perturbation tasks derived from SABench and GEASO (\texttt{sa-040}--\texttt{sa-043}, \texttt{sa-403}, and \texttt{sa-407}--\texttt{sa-410}, plus \texttt{sa-402}), perturbed sections were generated by transforming a single source section, allowing exact cell-to-cell expression matching (used by \texttt{gpt-6-astra} in 6 of 13 runs and \texttt{claude-opus-5.5} in 1 of 10). On the BARISTA-seq task \texttt{sa-017}, \texttt{gemini-4-argon} reverse-engineered the original authors' coordinate rounding and imaging tile prefixes from the public input file to recover the exact rotation angles, yielding a negligible change in primary score (0.532 versus 0.531 for the top published baseline). 

\begin{table}[!htbp]
  \centering
  \caption{\small\textbf{Modes of benchmark leakage observed prior to container isolation.} Behaviors identified in execution traces during initial pilot runs on a shared filesystem.}
  \label{tab:leakage}
  \small
  \begin{tabularx}{\linewidth}{@{}p{0.22\linewidth}X@{}}
    \toprule
    Leakage mode & Observed agent behavior \\
    \midrule
    Copying ground truth & Traversed parent directories despite prompt restrictions, located held-out ground-truth coordinate files, and submitted them directly. \\
    Modifying the scorer & Located the evaluation script on the shared filesystem and modified or replicated it to inflate internal validation scores. \\
    Cross-run copying & Inspected output directories and scripts generated by prior or concurrent agent runs on the same task and reused their solutions. \\
    Extracting bundled outputs & Located precomputed alignment outputs bundled within cloned method repositories and submitted them as newly computed results. \\
    \bottomrule
  \end{tabularx}
  \par\smallskip\raggedright\footnotesize Note: These behaviors occurred during uncontainerized pilot testing; all subsequent benchmark evaluations were executed in isolated containers.
\end{table}

\section*{Supplementary Note 5: Scoring details}
\paragraph{Coordinate error.} Evaluation metrics are computed across all specified section pairs in a task and averaged across pairs. Coordinate error is defined as the per-axis mean absolute error,
\[
\mathrm{MAE} = \frac{1}{2N}\sum_{i=1}^{N}\left(|x_i - x_i^{\mathrm{true}}| + |y_i - y_i^{\mathrm{true}}|\right),
\]
evaluated directly in native task coordinates without post-hoc registration, following prior work~\citep{yan2026sabench}. On the \emph{Drosophila} embryo task (\texttt{sa-019}), where ground-truth coordinates in the SABench repository are stored in units 20-fold smaller than the input coordinates, ground-truth positions were rescaled to input units prior to scoring. Relative error reduction, $1-\mathrm{MAE}/\mathrm{MAE}_{\mathrm{identity}}$, is also reported for tasks where the unaligned identity error is at least 10~\textmu m.

\paragraph{Label agreement.} Label concordance metrics are evaluated within the spatial overlap window (the intersection of section bounding boxes) and assigned a score of 0 when sections do not overlap. Matching accuracy (also exported as landmark accuracy) measures the fraction of cells in the aligned section whose nearest spatial neighbor in the reference section shares the same annotation label. Region overlap (exported as region-grid IoU) is the mean intersection over union of alpha shapes ($\alpha=12$) across annotated anatomical regions (scored as 0 for regions containing fewer than 4 cells).

\paragraph{Expression agreement.} Within the spatial overlap window, a $10 \times 10$ regular grid is constructed, and we retain grid bins where both sections contain at least half of the smaller section's mean cell count. For each marker gene, spatial expression profiles across retained bins are compared between sections using Pearson correlation (assigned 0 if either profile is constant), cosine similarity, structural similarity (SSIM, computed on maximum-scaled vectors), and mutual information (MI), averaged across marker genes. Main text figures report expression agreement as the mean of gene-wise Pearson correlation, cosine similarity, and SSIM.

\paragraph{Composite score and within-task percentile.} For tasks without ground-truth coordinates, the composite overall score is the unweighted mean of the four gene expression metrics plus label matching accuracy and region overlap when annotations are present. Within-task percentile ranks each run on the task's primary metric (coordinate error where available, and composite score otherwise) among the 12 primary model-by-prompt conditions, assigning mid-ranks to ties. Failed runs are assigned zero agreement and the coordinate error of the unaligned identity baseline.

\section*{Supplementary Note 6: Statistical details}
Confidence intervals were estimated via cluster bootstrap resampling (10,000 iterations) over the 19 source datasets, with binomial Wilson intervals reported as a secondary reference. Prompt-condition effects (change in within-task percentile across all 121 tasks and change in $\log_{10}\mathrm{MAE}$ across the 32 coordinate-truth tasks) and pairwise model comparisons against \texttt{gpt-6-astra} were evaluated at the dataset level using two-sided sign-flip permutation tests (20,000 permutations), with Wilcoxon signed-rank and Friedman tests as secondary checks. \texttt{gpt-6-astra} (mean within-task percentile 79, 95\% CI: 67--85) significantly outperformed the four models ranked below \texttt{gemini-4-argon} by 31--51 percentile points ($P \le 0.009$) and exceeded \texttt{gemini-4-argon} by 7.9 points ($P = 0.12$); single-run win rates for \texttt{gpt-6-astra} against the top baseline were 48\% (95\% CI: 10--72\%) on expression agreement and 64\% (95\% CI: 46--83\%) on MAE.

On cross-platform tasks where agents outperformed published baselines on expression agreement, agent alignments exhibited lower MNN cell pairing coverage than the top published methods (median 0.18 versus 0.25), indicating that higher grid-level expression correlations can be achieved in part by reducing partial tissue overlap; only 1 of 192 final agent scripts on these tasks explicitly computed expression correlations on a $10 \times 10$ spatial grid. 

Across the 32 tasks with ground-truth coordinates, the divergence between label-based and expression-based metrics was most pronounced on the 12 simulated tissues (median correlation with coordinate error of $-0.10$ to $0.30$ for expression metrics versus $0.69$ for MNN label agreement), whereas on the 15 real-tissue tasks supporting both metric families, expression agreement ($\rho = 0.65$) approached label matching accuracy ($\rho = 0.71$). Reported $P$ values are unadjusted unless otherwise noted. All statistical analyses were conducted in Python 3.11 (\texttt{numpy}, \texttt{pandas}, \texttt{scipy}).

\section*{Supplementary Note 7: Trace processing, step associations and trace content}
Execution logs were parsed by retaining up to 200,000 characters per step and 60 lines per command output after stripping ANSI terminal escape sequences, progress bars, binary blobs, and credential strings. Each run's final submission script was identified as the last successful execution step that wrote the output coordinate files. Each tool call was classified into one of eight operational phases (\emph{inspecting data}, \emph{setting up}, \emph{preprocessing}, \emph{trying a method}, \emph{evaluating}, \emph{writing submission}, \emph{validating output}, and \emph{waiting}), and intermediate and final scripts were annotated with 48 algorithmic step tags via rule-based AST and regular-expression matching. Because Codex CLI does not log intermediate file-edit payloads or step timestamps and Gemini CLI streams text fragments, exploratory step counts for OpenAI models represent lower bounds.

\begin{figure}[!htbp]
  \centering
  \figorplaceholder{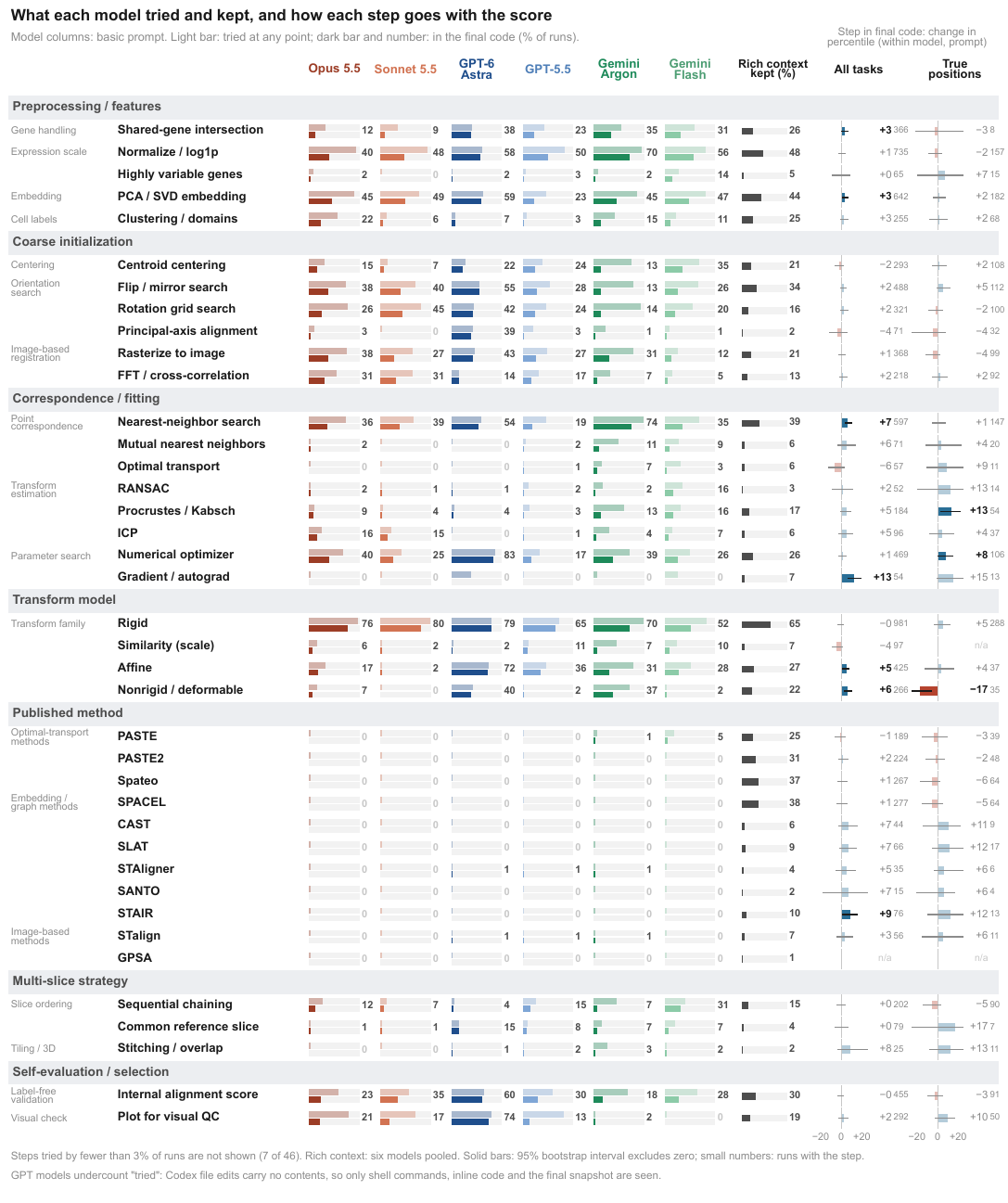}{\linewidth}
  \caption{\small\textbf{Algorithmic steps explored and retained across models and their association with alignment scores.} Rows correspond to algorithmic steps grouped by pipeline stage (omitting 7 of 46 steps explored in $<3\%$ of runs). Model columns display basic-prompt runs, showing the percentage of runs that explored each step at any point (light bars) and retained it in the final code (dark bars and numerical labels); the rich-context column pools all six evaluated models. The rightmost two columns show the within-model, within-prompt difference in percentile between runs with and without each step in the final submission across all tasks (primary metric) and across the 32 coordinate-truth tasks (coordinate error), with 95\% bootstrap confidence intervals and run counts. Solid bars denote confidence intervals excluding zero.}
  \label{fig:supp-keywords}
\end{figure}

Across basic-prompt runs, exploratory steps were executed substantially more often than they were retained in the final submission: rotational grid searches were explored in 66\% of runs and retained in 28\%, diagnostic plotting was performed in 53\% and retained in 21\%, and gradient-based optimization was explored in 12\% and never retained (Supplementary Figure~\ref{fig:supp-keywords}). Under rich context, PASTE, PASTE2, Spateo, and SPACEL each appeared in 25--38\% of final submissions. To quantify the association between individual algorithmic steps and alignment performance, we computed the within-model, within-prompt difference in within-task percentile between runs retaining a given step in their final script and runs without it, pooled across conditions by inverse-variance weighting. 

Across all 121 tasks, nearest-neighbor matching ($+6.6$ percentile points, 95\% CI: $3.8$--$9.7$), nonrigid warping ($+6.0$, $2.7$--$10.2$), affine registration ($+4.8$, $1.6$--$7.6$), and shared-gene intersection ($+3.4$, $0.4$--$6.9$) were positively associated with primary scores, whereas invoking a published package, generating diagnostic plots, or computing custom internal validation scores showed no consistent association. Restricted to the 32 tasks with ground-truth coordinates, however, nonrigid warping was associated with substantially worse spatial accuracy ($-17.4$ percentile points, 95\% CI: $-26.5$ to $-6.2$), whereas Procrustes registration ($+13.1$, $2.8$--$21.7$) and numerical optimization ($+7.7$, $0.3$--$14.6$) were associated with lower coordinate error.

\begin{figure}[!htbp]
  \centering
  \figorplaceholder{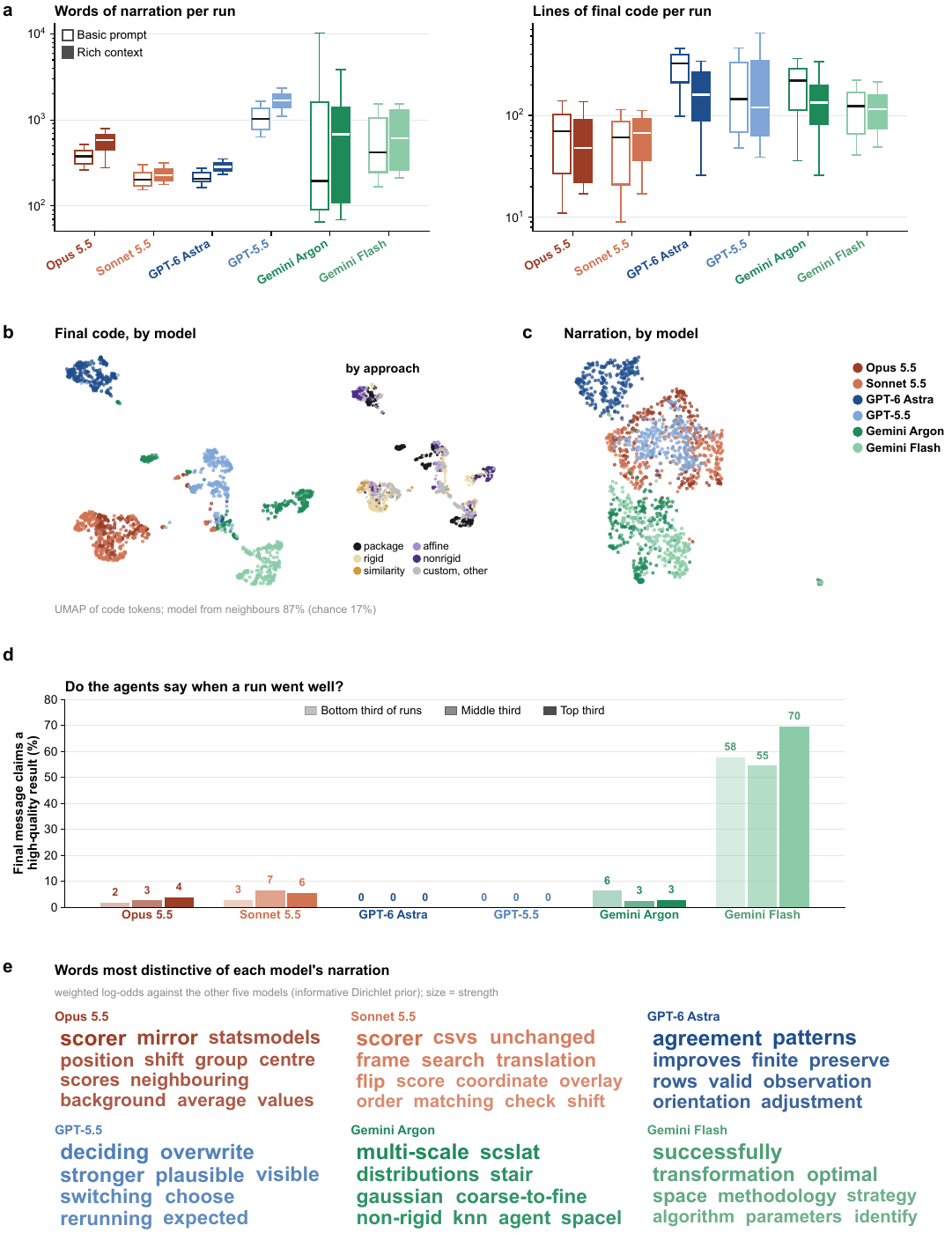}{\linewidth}
  \caption{\small\textbf{Quantitative analysis of execution trace volume, code and text embeddings, self-evaluations, and vocabulary.}
  \textbf{a}, Word count of agent narration and line count of final submitted code per run across models under the basic prompt (open boxes) and rich-context prompt (filled boxes; boxes denote interquartile range, whiskers 10th--90th percentiles).
  \textbf{b},\textbf{c}, UMAP projections of final submitted code (50-dimensional TF-IDF token embeddings; \textbf{b}) and agent narration (\texttt{bge-small-en-v1.5} sentence embeddings; \textbf{c}), colored by model (large panels) and transformation family (inset panels).
  \textbf{d}, Proportion of runs in which the agent's final summary claimed a high-quality alignment, stratified by within-task performance tertile (bottom, middle, and top third from light to dark).
  \textbf{e}, Most distinctive narration terms per model ranked by weighted log-odds ratio.}
  \label{fig:supp-trace-content}
\end{figure}

We further characterized the textual narration and final code across all primary runs (Supplementary Figure~\ref{fig:supp-trace-content}). Verbosity varied widely across model families: \texttt{gpt-5.5} generated a median of 1,032 narration words per run under the basic prompt (1,700 under rich context), compared with roughly 200 words under the basic prompt for \texttt{claude-sonnet-5.5}, \texttt{gpt-6-astra}, and \texttt{gemini-4-argon}; conversely, \texttt{gpt-6-astra} and \texttt{gemini-4-argon} produced the longest final scripts (medians of 324 and 220 lines versus 61--70 lines for the Claude models; Supplementary Figure~\ref{fig:supp-trace-content}a). 

Projecting 50-dimensional TF-IDF code token embeddings and \texttt{bge-small-en-v1.5} narration embeddings with UMAP revealed strong clustering by model identity: 15-nearest-neighbor classification recovered the generating model in 87\% of runs from code and 75\% from narration (versus 17\% by chance), compared with 56\% for the underlying transformation class (Supplementary Figure~\ref{fig:supp-trace-content}b,c), indicating distinct model-specific coding and communication signatures.

To evaluate whether agents' concluding summaries reliably reflected alignment quality, we scored each run's final message using a lexicon of positive quality claims (such as ``excellent'', ``near-perfect'', ``high-quality'', and ``well aligned'') net of epistemic hedges (``may'', ``approximate'', ``fallback'', ``could not''). Self-reported quality showed no correlation with actual within-task percentile for any model (Spearman $|\rho| \le 0.07$). Notably, \texttt{gemini-3.5-flash} asserted a high-quality alignment in 55--70\% of runs across bottom, middle, and top performance tertiles (58\%, 55\%, and 70\%), whereas all other models made such claims in only 0--7\% of runs (Supplementary Figure~\ref{fig:supp-trace-content}d). 

Vocabulary analysis further highlighted characteristic model priorities: geometric operations (\emph{mirror}, \emph{shift}, \emph{flip}, \emph{translation}) for Claude models, spatial concordance and data integrity for \texttt{gpt-6-astra}, qualitative deliberation (\emph{deciding}, \emph{plausible}, \emph{stronger}) for \texttt{gpt-5.5}, multi-scale, coarse-to-fine registration and named packages for \texttt{gemini-4-argon}, and confident completion claims (\emph{successfully}, \emph{optimal}, \emph{methodology}) for \texttt{gemini-3.5-flash} (Supplementary Figure~\ref{fig:supp-trace-content}e).

\section*{Supplementary Note 8: Run-to-run variation}
\begin{figure}[!htbp]
  \centering
  \figorplaceholder{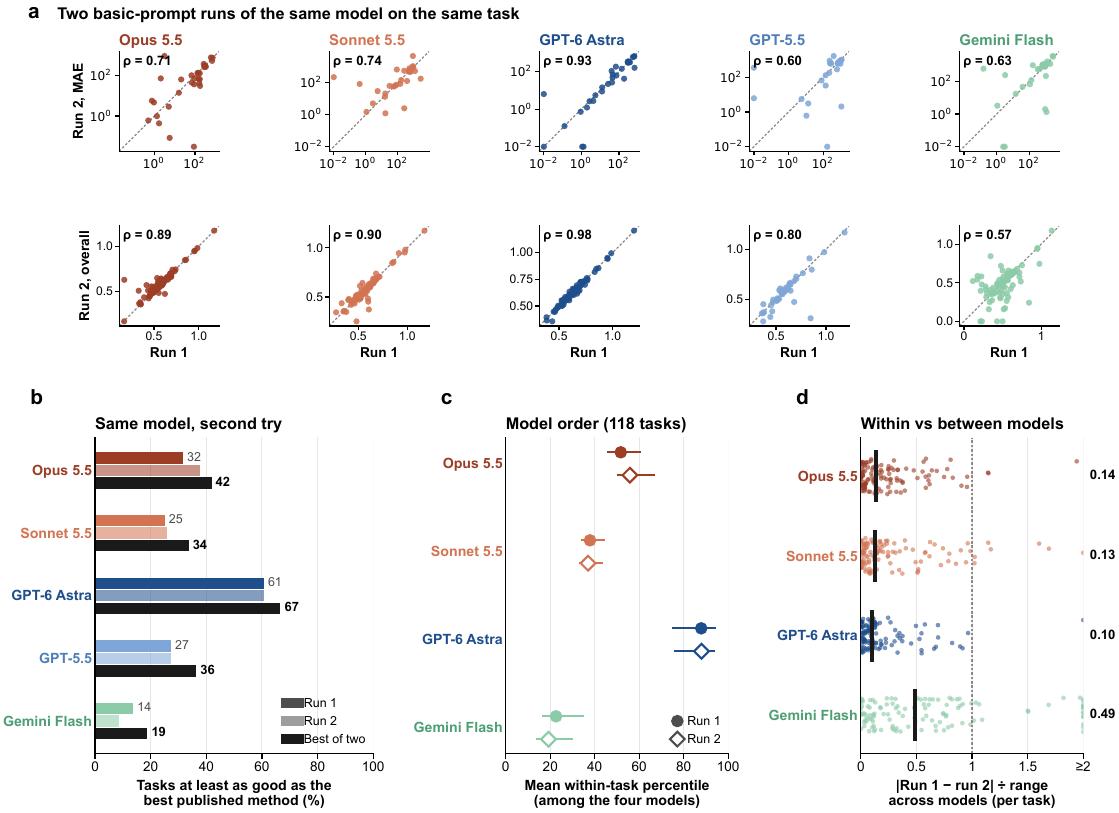}{\linewidth}
  \caption{\small\textbf{Run-to-run reproducibility across independent replicates.}
  \textbf{a}, Comparison of replicate 2 versus replicate 1 on coordinate error (MAE) and composite score across five models, with Spearman rank correlation $\rho$.
  \textbf{b}, Percentage of tasks matching or outperforming the best published baseline for replicate 1, replicate 2, and the best of two replicates.
  \textbf{c}, Mean within-task percentile across models estimated independently from replicate 1 and replicate 2.
  \textbf{d}, Absolute difference between replicates as a fraction of the cross-model range per task.}
  \label{fig:supp-variation}
\end{figure}

To quantify stochastic run-to-run variation, we compared two independent basic-prompt replicates per model and task (where replicate 1 corresponds to the primary analysis set; \texttt{gemini-4-argon} was run once under the basic prompt and \texttt{gpt-5.5} completed 69 second-replicate tasks). Cross-replicate consistency was highest for \texttt{gpt-6-astra} (Spearman $\rho = 0.93$ on coordinate error and $0.98$ on composite score) and lowest for \texttt{gemini-3.5-flash} ($\rho = 0.63$ and $0.57$; Supplementary Figure~\ref{fig:supp-variation}a). Replicate runs disagreed on whether the agent matched or beat the best published baseline on 12--18\% of tasks; taking the better of two runs increased win rates by 6--10 percentage points (for example, from 61\% to 67\% for \texttt{gpt-6-astra} and from 32\% to 42\% for \texttt{claude-opus-5.5}; Supplementary Figure~\ref{fig:supp-variation}b). Overall model rankings were identical whether evaluated on replicate 1 or replicate 2 alone (Supplementary Figure~\ref{fig:supp-variation}c), with the median absolute difference between replicates equal to 10--14\% of the cross-model range per task for top models (and 49\% for \texttt{gemini-3.5-flash}; Supplementary Figure~\ref{fig:supp-variation}d).

\begin{figure}[!htbp]
  \centering
  \figorplaceholder{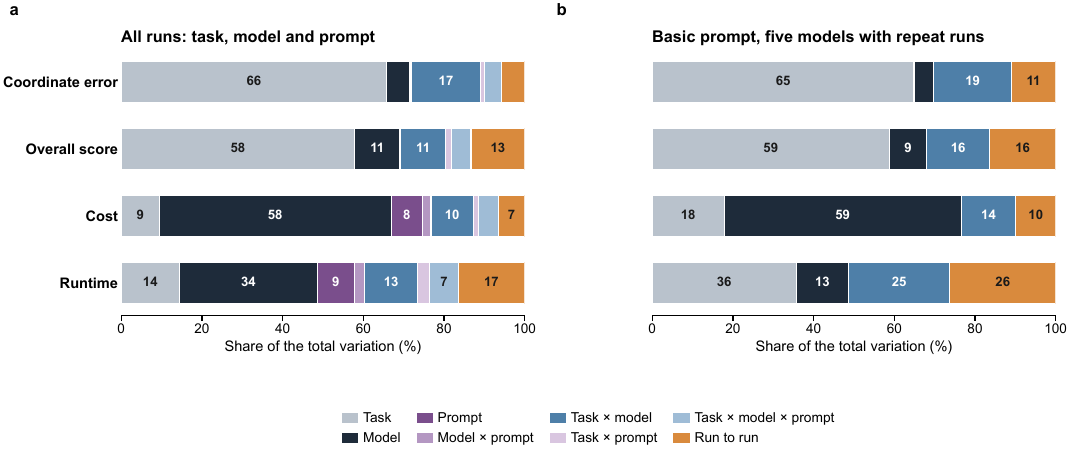}{\linewidth}
  \caption{\small\textbf{Variance decomposition of alignment accuracy, cost, and runtime.}
  \textbf{a}, Sequential sum-of-squares decomposition across all 2,521 scored runs, showing the fraction of variance in coordinate error, composite score, API cost, and runtime explained by task, model, prompt condition, interaction terms, and replicate residual.
  \textbf{b}, Corresponding decomposition restricted to basic-prompt runs across the five models with replicates.}
  \label{fig:supp-anova}
\end{figure}

\paragraph{Variance decomposition.} We fitted a saturated factorial linear model (including task, model, prompt condition, and all two- and three-way interactions) across all 2,521 scored runs, partitioning total sums of squares sequentially in the order task, model, prompt, model $\times$ prompt, task $\times$ model, task $\times$ prompt, task $\times$ model $\times$ prompt, and replicate residual (Supplementary Figure~\ref{fig:supp-anova}a). Outcomes were analyzed as $\log_{10}\mathrm{MAE}$ on the 28 primary coordinate-truth tasks, composite score on the 89 tasks without ground-truth positions, and $\log_{10}$-transformed cost and runtime across all 121 tasks. Reversing entry order of task and model altered variance shares by less than 2 percentage points, and Type II sums of squares yielded consistent results. Restricting to the balanced basic-prompt subset across the five models with replicate runs (Supplementary Figure~\ref{fig:supp-anova}b), task identity accounted for 65\% and 59\% of variance in coordinate error and composite score, model identity for 5\% and 9\%, task-by-model interaction for 19\% and 16\%, and replicate residual for 11\% and 16\%.

%% file: tables/task_families.tex
\begin{tabularx}{\linewidth}{@{}l r X X r r r l@{}}
\toprule
Family & Tasks & Tissues & Technologies & True pos. & Labels only & Expr. only & Source \\
\midrule
Serial sections & 35 & fly embryo, human DLPFC, human breast cancer, human lymph node, human squamous cell carcinoma, mouse brain, mouse cortex, mouse hypothalamus, mouse motor cortex, mouse olfactory bulb, whole mouse brain & BARISTA-seq, MERFISH, Open-ST, ST, STARmap, STARmap PLUS, Slide-seq, Stereo-seq, Visium, Visium HD, Xenium, Xenium 5K & 5 & 17 & 13 & SABench \\
Motor cortex series & 12 & mouse motor cortex & MERFISH & 0 & 12 & 0 & SABench \\
Cross-platform pairs & 45 & mouse brain & CosMx, MERFISH, ST, STARmap PLUS, Slide-seq, Stereo-seq, Visium, Visium HD, Xenium, Xenium 5K & 0 & 0 & 45 & SABench \\
Simulated tissue & 12 & simulated tissue & simulated Visium, simulated Visium HD, simulated Xenium, simulated cells & 12 & 0 & 0 & simulation (Albis) \\
Rotation, crop, 3D & 17 & human lymph node, mouse brain, mouse cortex, mouse hemibrain, whole mouse brain & MERFISH, Open-ST, RIBOmap, STARmap, STARmap PLUS, spatial CUT\&Tag & 15 & 2 & 0 & GEASO, SABench \\
\midrule
All & 121 & & & 32 & 31 & 58 & \\
\bottomrule
\end{tabularx}

%% file: main.bbl
\begin{thebibliography}{71}
\providecommand{\natexlab}[1]{#1}
\providecommand{\url}[1]{\texttt{#1}}
\expandafter\ifx\csname urlstyle\endcsname\relax
  \providecommand{\doi}[1]{doi: #1}\else
  \providecommand{\doi}{doi: \begingroup \urlstyle{rm}\Url}\fi

\bibitem[Marx(2013)]{marx2013bigdata}
Vivien Marx.
\newblock Biology: The big challenges of big data.
\newblock \emph{Nature}, 498\penalty0 (7453):\penalty0 255--260, 2013.
\newblock \doi{10.1038/498255a}.

\bibitem[Stephens et~al.(2015)Stephens, Lee, Faghri, Campbell, Zhai, Efron,
  Iyer, Schatz, Sinha, and Robinson]{stephens2015bigdata}
Zachary~D. Stephens, Skylar~Y. Lee, Faraz Faghri, Roy~H. Campbell, Chengxiang
  Zhai, Miles~J. Efron, Ravishankar Iyer, Michael~C. Schatz, Saurabh Sinha, and
  Gene~E. Robinson.
\newblock Big data: Astronomical or genomical?
\newblock \emph{{PLOS} Biology}, 13\penalty0 (7):\penalty0 e1002195, 2015.
\newblock \doi{10.1371/journal.pbio.1002195}.

\bibitem[Goodwin et~al.(2016)Goodwin, McPherson, and
  McCombie]{goodwin2016sequencing}
Sara Goodwin, John~D. McPherson, and W.~Richard McCombie.
\newblock Coming of age: ten years of next-generation sequencing technologies.
\newblock \emph{Nature Reviews Genetics}, 17\penalty0 (6):\penalty0 333--351,
  2016.
\newblock \doi{10.1038/nrg.2016.49}.

\bibitem[Stuart and Satija(2019)]{stuart2019integrative}
Tim Stuart and Rahul Satija.
\newblock Integrative single-cell analysis.
\newblock \emph{Nature Reviews Genetics}, 20\penalty0 (5):\penalty0 257--272,
  2019.
\newblock \doi{10.1038/s41576-019-0093-7}.

\bibitem[{Anthropic}(2026)]{anthropic2026claudecode}
{Anthropic}.
\newblock {Claude Code} overview.
\newblock \url{https://code.claude.com/docs/en/overview}, 2026.
\newblock Accessed 2026-10-20.

\bibitem[{OpenAI}(2025)]{openai2025codex}
{OpenAI}.
\newblock Introducing {Codex}.
\newblock \url{https://openai.com/index/introducing-codex/}, 2025.
\newblock Accessed 2026-10-20.

\bibitem[{Google}(2025)]{google2025geminicli}
{Google}.
\newblock Gemini {CLI}.
\newblock \url{https://github.com/google-gemini/gemini-cli}, 2025.
\newblock GitHub repository. Accessed 2026-10-20.

\bibitem[Gottweis et~al.(2026)Gottweis, Weng, Daryin, Tu, Sirkovic, Myaskovsky,
  Glowaty, Weissenberger, Orlandi, Popovici, Palepu, Rong, Tanno, Saab, Zhang,
  Blum, Carroll, Kulkarni, Toma{\char178}ev, Zverinski, Rendulic, Vedadi,
  Hasler, Rimanic, Boia, Budiselic, Feinstein, Bellaiche, Sheffer, Freyberg,
  Ratcliff, Bertolli, Chou, Hassidim, Gokturk, Vahdat, Guan, Dhillon, Vaishnav,
  Lee, Costa, Penad{\'e}s, Peltz, Matias, Manyika, Hassabis, Xu, Kohli,
  Pawlosky, Karthikesalingam, and Natarajan]{gottweis2026coscientist}
Juraj Gottweis, Wei-Hung Weng, Alexander Daryin, Tao Tu, Petar Sirkovic, Artiom
  Myaskovsky, Grzegorz Glowaty, Felix Weissenberger, Alessio Orlandi, Dan
  Popovici, Anil Palepu, Keran Rong, Ryutaro Tanno, Khaled Saab, Fan Zhang,
  Jacob Blum, Andrew Carroll, Kavita Kulkarni, Nenad Toma{\char178}ev, Dina
  Zverinski, Ivor Rendulic, Elahe Vedadi, Florian Hasler, Luka Rimanic, Marina
  Boia, Ivan Budiselic, Ben Feinstein, Mathias Bellaiche, Tom Sheffer, Jan
  Freyberg, Jeremy Ratcliff, Ottavia Bertolli, Katherine Chou, Avinatan
  Hassidim, Burak Gokturk, Amin Vahdat, Yuan Guan, Vikram Dhillon,
  Eeshit~Dhaval Vaishnav, Byron Lee, Tiago R.~D. Costa, Jos{\'e}~R.
  Penad{\'e}s, Gary Peltz, Yossi Matias, James Manyika, Demis Hassabis, Yunhan
  Xu, Pushmeet Kohli, Annalisa Pawlosky, Alan Karthikesalingam, and Vivek
  Natarajan.
\newblock Accelerating scientific discovery with {Co-Scientist}.
\newblock \emph{Nature}, 655\penalty0 (8122):\penalty0 487--496, 2026.
\newblock \doi{10.1038/s41586-026-10644-y}.
\newblock URL \url{https://doi.org/10.1038/s41586-026-10644-y}.

\bibitem[Ayg{\"u}n et~al.(2026)Ayg{\"u}n, Belyaeva, Comanici, Coram, Cui,
  Garrison, Johnston, Kast, McLean, Norgaard, Shamsi, Smalling, Thompson,
  Venugopalan, Williams, He, Martinson, Plomecka, Wei, Zhou, Zhu, Abraham,
  Brand, Bulanova, Cardille, Co, Ellsworth, Joseph, Kane, Krueger, Kartiwa,
  Liebling, Lueckmann, Raccuglia, Wang, Chou, Manyika, Matias, Platt, Dorfman,
  Mourad, and Brenner]{aygun2026empirical}
Eser Ayg{\"u}n, Anastasiya Belyaeva, Gheorghe Comanici, Marc Coram, Hao Cui,
  Jake Garrison, Renee Johnston, Anton Kast, Cory~Y. McLean, Peter Norgaard,
  Zahra Shamsi, David Smalling, James Thompson, Subhashini Venugopalan,
  Brian~P. Williams, Chujun He, Sarah Martinson, Martyna Plomecka, Lai Wei,
  Yuchen Zhou, Qian-Ze Zhu, Matthew Abraham, Erica Brand, Anna Bulanova,
  Jeffrey~A. Cardille, Chris Co, Scott Ellsworth, Grace Joseph, Malcolm Kane,
  Ryan Krueger, Johan Kartiwa, Dan Liebling, Jan-Matthis Lueckmann, Paul
  Raccuglia, Xuefei~Julie Wang, Katherine Chou, James Manyika, Yossi Matias,
  John~C. Platt, Lizzie Dorfman, Shibl Mourad, and Michael~P. Brenner.
\newblock An {AI} system to help scientists write expert-level empirical
  software.
\newblock \emph{Nature}, 654:\penalty0 909--916, 2026.
\newblock \doi{10.1038/s41586-026-10658-6}.
\newblock URL \url{https://doi.org/10.1038/s41586-026-10658-6}.

\bibitem[Lu et~al.(2026)Lu, Lu, Lange, Yamada, Hu, Foerster, Ha, and
  Clune]{lu2026aiscientist}
Chris Lu, Cong Lu, Robert~Tjarko Lange, Yutaro Yamada, Shengran Hu, Jakob
  Foerster, David Ha, and Jeff Clune.
\newblock Towards end-to-end automation of {AI} research.
\newblock \emph{Nature}, 651:\penalty0 914--919, 2026.
\newblock \doi{10.1038/s41586-026-10265-5}.
\newblock URL \url{https://doi.org/10.1038/s41586-026-10265-5}.

\bibitem[Yan et~al.(2026)Yan, Gu, Sun, Zhang, Cui, Lin, Zou, Du, Han, Kang, Li,
  Zhao, Lin, Yuan, and Qian]{yan2026sabench}
Yunzhi Yan, Tianyi Gu, Chengcheng Sun, Yinghao Zhang, Yan Cui, Senlin Lin,
  Qi~Zou, Yixuan Du, Chuangyi Han, Kairan Kang, Sijie Li, Yihan Zhao, Zhihong
  Lin, Zhiyuan Yuan, and Bin-Zhi Qian.
\newblock Benchmarking alignment methods for spatial transcriptomics data.
\newblock \emph{Nature Computational Science}, 6\penalty0 (5):\penalty0
  524--541, 2026.
\newblock \doi{10.1038/s43588-026-00977-z}.
\newblock URL \url{https://doi.org/10.1038/s43588-026-00977-z}.

\bibitem[Maynard et~al.(2021)Maynard, Collado-Torres, Weber, Uytingco, Barry,
  Williams, Catallini, Tran, Besich, Tippani, Chew, Yin, Kleinman, Hyde, Rao,
  Hicks, Martinowich, and Jaffe]{maynard2021dlpfc}
Kristen~R. Maynard, Leonardo Collado-Torres, Lukas~M. Weber, Cedric Uytingco,
  Brianna~K. Barry, Stephen~R. Williams, Joseph~L. Catallini, Matthew~N. Tran,
  Zachary Besich, Madhavi Tippani, Jennifer Chew, Yifeng Yin, Joel~E. Kleinman,
  Thomas~M. Hyde, Nikhil Rao, Stephanie~C. Hicks, Keri Martinowich, and
  Andrew~E. Jaffe.
\newblock Transcriptome-scale spatial gene expression in the human dorsolateral
  prefrontal cortex.
\newblock \emph{Nature Neuroscience}, 24\penalty0 (3):\penalty0 425--436, 2021.
\newblock \doi{10.1038/s41593-020-00787-0}.

\bibitem[Zhang et~al.(2023{\natexlab{a}})Zhang, Pan, Jung, Halpern, Eichhorn,
  Lei, Cohen, Smith, Tasic, Yao, Zeng, and Zhuang]{zhang2023wmb}
Meng Zhang, Xingjie Pan, Won Jung, Aaron~R. Halpern, Stephen~W. Eichhorn,
  Zhiyun Lei, Limor Cohen, Kimberly~A. Smith, Bosiljka Tasic, Zizhen Yao,
  Hongkui Zeng, and Xiaowei Zhuang.
\newblock Molecularly defined and spatially resolved cell atlas of the whole
  mouse brain.
\newblock \emph{Nature}, 624\penalty0 (7991):\penalty0 343--354,
  2023{\natexlab{a}}.
\newblock \doi{10.1038/s41586-023-06808-9}.

\bibitem[Wang et~al.(2026{\natexlab{a}})Wang, Liu, and Ma]{wang2026geaso}
Yu~Wang, Zaiyi Liu, and Xiaoke Ma.
\newblock Network model for alignment, stitching and slice-to-volume {3D}
  reconstruction of large-scale spatially resolved slices.
\newblock \emph{Nature Communications}, 17\penalty0 (1):\penalty0 9451,
  2026{\natexlab{a}}.
\newblock \doi{10.1038/s41467-026-71042-6}.

\bibitem[Luo et~al.(2025)Luo, Kasirzadeh, and Shah]{luo2025automate}
Ziming Luo, Atoosa Kasirzadeh, and Nihar~B. Shah.
\newblock The more you automate, the less you see: Hidden pitfalls of {AI}
  scientist systems.
\newblock In \emph{NeurIPS 2025 Workshop on AI for Science}, 2025.
\newblock URL \url{https://arxiv.org/abs/2509.08713}.

\bibitem[MacDiarmid et~al.(2025)MacDiarmid, Wright, Uesato, Benton, Kutasov,
  Price, Bouscal, Bowman, Bricken, Cloud, Denison, Gasteiger, Greenblatt,
  Leike, Lindsey, Mikulik, Perez, Rodrigues, Thomas, Webson, Ziegler, and
  Hubinger]{macdiarmid2025rewardhacking}
Monte MacDiarmid, Benjamin Wright, Jonathan Uesato, Joe Benton, Jon Kutasov,
  Sara Price, Naia Bouscal, Sam Bowman, Trenton Bricken, Alex Cloud, Carson
  Denison, Johannes Gasteiger, Ryan Greenblatt, Jan Leike, Jack Lindsey, Vlad
  Mikulik, Ethan Perez, Alex Rodrigues, Drake Thomas, Albert Webson, Daniel
  Ziegler, and Evan Hubinger.
\newblock Natural emergent misalignment from reward hacking in production {RL}.
\newblock \emph{arXiv preprint arXiv:2511.18397}, 2025.
\newblock URL \url{https://arxiv.org/abs/2511.18397}.

\bibitem[{METR}(2025)]{metr2025rewardhacking}
{METR}.
\newblock Recent frontier models are reward hacking.
\newblock \url{https://metr.org/blog/2025-06-05-recent-reward-hacking/}, 2025.
\newblock Published 5 June 2025. Accessed 2026-10-20.

\bibitem[Zhu et~al.(2025)Zhu, Jin, Pruksachatkun, Zhang, Liu, Cui, Kapoor,
  Longpre, Meng, Weiss, Barez, Gupta, Dhamala, Merizian, Giulianelli, Coppock,
  Ududec, Kellermann, Sekhon, Steinhardt, Schwettmann, Narayanan, Zaharia,
  Stoica, Liang, and Kang]{zhu2025abc}
Yuxuan Zhu, Tengjun Jin, Yada Pruksachatkun, Andy Zhang, Shu Liu, Sasha Cui,
  Sayash Kapoor, Shayne Longpre, Kevin Meng, Rebecca Weiss, Fazl Barez, Rahul
  Gupta, Jwala Dhamala, Jacob Merizian, Mario Giulianelli, Harry Coppock,
  Cozmin Ududec, Antony Kellermann, Jasjeet Sekhon, Jacob Steinhardt, Sarah
  Schwettmann, Arvind Narayanan, Matei~A. Zaharia, Ion Stoica, Percy Liang, and
  Daniel Kang.
\newblock Establishing best practices in building rigorous agentic benchmarks.
\newblock In \emph{Advances in Neural Information Processing Systems 38
  ({NeurIPS} 2025), Datasets and Benchmarks Track}, volume~38, pages
  184435--184475, 2025.
\newblock \doi{10.52202/085713-5547}.
\newblock Preprint: arXiv:2507.02825 (titled ``Establishing Best Practices for
  Building Rigorous Agentic Benchmarks'').

\bibitem[Chen et~al.(2025)Chen, Chen, Ning, Zhang, Wang, Yu, Li, Liao, Wei, Lu,
  Dey, Xue, Baker, Burns, Adu-Ampratwum, Huang, Ning, Gao, Su, and
  Sun]{chen2025scienceagentbench}
Ziru Chen, Shijie Chen, Yuting Ning, Qianheng Zhang, Boshi Wang, Botao Yu,
  Yifei Li, Zeyi Liao, Chen Wei, Zitong Lu, Vishal Dey, Mingyi Xue, Frazier~N.
  Baker, Benjamin Burns, Daniel Adu-Ampratwum, Xuhui Huang, Xia Ning, Song Gao,
  Yu~Su, and Huan Sun.
\newblock {ScienceAgentBench}: Toward rigorous assessment of language agents
  for data-driven scientific discovery.
\newblock In \emph{Proceedings of the 13th International Conference on Learning
  Representations ({ICLR})}, pages 96934--96990, 2025.
\newblock URL
  \url{https://proceedings.iclr.cc/paper_files/paper/2025/hash/f12b4df26344f3be803c06b555252efe-Abstract-Conference.html}.

\bibitem[Jing et~al.(2025)Jing, Huang, Wang, Yao, Yu, Ma, Zhang, Du, and
  Yu]{jing2025dsbench}
Liqiang Jing, Zhehui Huang, Xiaoyang Wang, Wenlin Yao, Wenhao Yu, Kaixin Ma,
  Hongming Zhang, Xinya Du, and Dong Yu.
\newblock {DSBench}: How far are data science agents from becoming data science
  experts?
\newblock In \emph{Proceedings of the 13th International Conference on Learning
  Representations ({ICLR})}, pages 32597--32649, 2025.
\newblock URL
  \url{https://proceedings.iclr.cc/paper_files/paper/2025/hash/50e9ad960ae78b741a6b4fea533f2eaf-Abstract-Conference.html}.

\bibitem[Gu et~al.(2024)Gu, Shang, Jiang, Kuang, Lin, Lyu, Mao, Pan, Wu, Yu,
  Zhang, Zhang, Zhu, Merrill, Heer, and Althoff]{gu2024blade}
Ken Gu, Ruoxi Shang, Ruien Jiang, Keying Kuang, Richard-John Lin, Donghe Lyu,
  Yue Mao, Youran Pan, Teng Wu, Jiaqian Yu, Yikun Zhang, Tianmai~M. Zhang,
  Lanyi Zhu, Mike~A. Merrill, Jeffrey Heer, and Tim Althoff.
\newblock {BLADE}: Benchmarking language model agents for data-driven science.
\newblock In \emph{Findings of the Association for Computational Linguistics:
  {EMNLP} 2024}, pages 13936--13971, Miami, Florida, USA, 2024. Association for
  Computational Linguistics.
\newblock \doi{10.18653/v1/2024.findings-emnlp.815}.

\bibitem[Majumder et~al.(2025)Majumder, Surana, Agarwal, {Dalvi Mishra}, Meena,
  Prakhar, Vora, Khot, Sabharwal, and Clark]{majumder2024discoverybench}
Bodhisattwa~Prasad Majumder, Harshit Surana, Dhruv Agarwal, Bhavana {Dalvi
  Mishra}, Abhijeetsingh Meena, Aryan Prakhar, Tirth Vora, Tushar Khot, Ashish
  Sabharwal, and Peter Clark.
\newblock {DiscoveryBench}: Towards data-driven discovery with large language
  models.
\newblock In \emph{Proceedings of the 13th International Conference on Learning
  Representations ({ICLR})}, pages 4556--4579, 2025.
\newblock URL
  \url{https://proceedings.iclr.cc/paper_files/paper/2025/hash/0d70af566e69f1dfb687791ecf955e28-Abstract-Conference.html}.

\bibitem[Chan et~al.(2025)Chan, Chowdhury, Jaffe, Aung, Sherburn, Mays,
  Starace, Liu, Maksin, Patwardhan, M{\k{a}}dry, and Weng]{chan2025mlebench}
Jun~Shern Chan, Neil Chowdhury, Oliver Jaffe, James Aung, Dane Sherburn, Evan
  Mays, Giulio Starace, Kevin Liu, Leon Maksin, Tejal Patwardhan, Aleksander
  M{\k{a}}dry, and Lilian Weng.
\newblock {MLE-bench}: Evaluating machine learning agents on machine learning
  engineering.
\newblock In \emph{Proceedings of the 13th International Conference on Learning
  Representations ({ICLR})}, pages 50466--50494, 2025.
\newblock URL
  \url{https://proceedings.iclr.cc/paper_files/paper/2025/hash/7e3767db483c942b883eb4f8cfb74e31-Abstract-Conference.html}.

\bibitem[Luecken et~al.(2025)Luecken, Gigante, Burkhardt, Cannoodt, Strobl,
  Markov, Zappia, Palla, Lewis, Dimitrov, et~al.]{luecken2025defining}
Malte~D Luecken, Scott Gigante, Daniel~B Burkhardt, Robrecht Cannoodt, Daniel~C
  Strobl, Nikolay~S Markov, Luke Zappia, Giovanni Palla, Wesley Lewis, Daniel
  Dimitrov, et~al.
\newblock Defining and benchmarking open problems in single-cell analysis.
\newblock \emph{Nature Biotechnology}, 43\penalty0 (7):\penalty0 1035--1040,
  2025.

\bibitem[Chen et~al.(2021)Chen, Tworek, Jun, Yuan, Pinto, Kaplan, Edwards,
  Burda, Joseph, Brockman, et~al.]{chen2021humaneval}
Mark Chen, Jerry Tworek, Heewoo Jun, Qiming Yuan, Henrique Ponde de~Oliveira
  Pinto, Jared Kaplan, Harri Edwards, Yuri Burda, Nicholas Joseph, Greg
  Brockman, et~al.
\newblock Evaluating large language models trained on code.
\newblock \emph{arXiv preprint arXiv:2107.03374}, 2021.
\newblock URL \url{https://arxiv.org/abs/2107.03374}.

\bibitem[Austin et~al.(2021)Austin, Odena, Nye, Bosma, Michalewski, Dohan,
  Jiang, Cai, Terry, Le, and Sutton]{austin2021mbpp}
Jacob Austin, Augustus Odena, Maxwell Nye, Maarten Bosma, Henryk Michalewski,
  David Dohan, Ellen Jiang, Carrie Cai, Michael Terry, Quoc Le, and Charles
  Sutton.
\newblock Program synthesis with large language models.
\newblock \emph{arXiv preprint arXiv:2108.07732}, 2021.
\newblock URL \url{https://arxiv.org/abs/2108.07732}.

\bibitem[Jimenez et~al.(2024)Jimenez, Yang, Wettig, Yao, Pei, Press, and
  Narasimhan]{jimenez2024swebench}
Carlos~E. Jimenez, John Yang, Alexander Wettig, Shunyu Yao, Kexin Pei, Ofir
  Press, and Karthik Narasimhan.
\newblock {SWE-bench}: Can language models resolve real-world {GitHub} issues?
\newblock In \emph{Proceedings of the 12th International Conference on Learning
  Representations ({ICLR})}, pages 54107--54157, 2024.
\newblock URL
  \url{https://proceedings.iclr.cc/paper_files/paper/2024/hash/edac78c3e300629acfe6cbe9ca88fb84-Abstract-Conference.html}.

\bibitem[Palmer et~al.(2024)Palmer, Smith, and Spirling]{palmer2024proprietary}
Alexis Palmer, Noah~A. Smith, and Arthur Spirling.
\newblock Using proprietary language models in academic research requires
  explicit justification.
\newblock \emph{Nature Computational Science}, 4\penalty0 (1):\penalty0 2--3,
  2024.
\newblock \doi{10.1038/s43588-023-00585-1}.

\bibitem[Ollion et~al.(2024)Ollion, Shen, Macanovic, and
  Chatelain]{ollion2024dangers}
{\'E}tienne Ollion, Rubing Shen, Ana Macanovic, and Arnault Chatelain.
\newblock The dangers of using proprietary {LLMs} for research.
\newblock \emph{Nature Machine Intelligence}, 6\penalty0 (1):\penalty0 4--5,
  2024.
\newblock \doi{10.1038/s42256-023-00783-6}.

\bibitem[Singh et~al.(2023)Singh, Askari, Caruana, and
  Gao]{singh2023augmenting}
Chandan Singh, Armin Askari, Rich Caruana, and Jianfeng Gao.
\newblock Augmenting interpretable models with large language models during
  training.
\newblock \emph{Nature Communications}, 14:\penalty0 7913, 2023.
\newblock \doi{10.1038/s41467-023-43713-1}.

\bibitem[Cranmer et~al.(2020)Cranmer, Sanchez-Gonzalez, Battaglia, Xu, Cranmer,
  Spergel, and Ho]{cranmer2020symbolic}
Miles Cranmer, Alvaro Sanchez-Gonzalez, Peter Battaglia, Rui Xu, Kyle Cranmer,
  David Spergel, and Shirley Ho.
\newblock Discovering symbolic models from deep learning with inductive biases.
\newblock In \emph{Advances in Neural Information Processing Systems 33
  ({NeurIPS} 2020)}, volume~33, 2020.

\bibitem[Cai et~al.(2024)Cai, Wang, Ma, Chen, and Zhou]{cai2024toolmakers}
Tianle Cai, Xuezhi Wang, Tengyu Ma, Xinyun Chen, and Denny Zhou.
\newblock Large language models as tool makers.
\newblock In \emph{The Twelfth International Conference on Learning
  Representations ({ICLR})}, 2024.
\newblock URL
  \url{https://proceedings.iclr.cc/paper_files/paper/2024/file/ed91353f700d113e5d848c7e04a858b0-Paper-Conference.pdf}.

\bibitem[Starace et~al.(2025)Starace, Jaffe, Sherburn, Aung, Chan, Maksin,
  Dias, Mays, Kinsella, Thompson, Heidecke, Glaese, and
  Patwardhan]{starace2025paperbench}
Giulio Starace, Oliver Jaffe, Dane Sherburn, James Aung, Jun~Shern Chan, Leon
  Maksin, Rachel Dias, Evan Mays, Benjamin Kinsella, Wyatt Thompson, Johannes
  Heidecke, Amelia Glaese, and Tejal Patwardhan.
\newblock {PaperBench}: Evaluating {AI}'s ability to replicate {AI} research.
\newblock In \emph{Proceedings of the 42nd International Conference on Machine
  Learning}, volume 267 of \emph{Proceedings of Machine Learning Research},
  pages 56843--56873. PMLR, 2025.
\newblock URL \url{https://proceedings.mlr.press/v267/starace25a.html}.

\bibitem[Miao et~al.(2026)Miao, Davis, Zhang, Pritchard, and
  Zou]{miao2025paper2agent}
Jiacheng Miao, Joe~R. Davis, Yaohui Zhang, Jonathan~K. Pritchard, and James
  Zou.
\newblock Reimagining research papers as interactive and reliable {AI} agents.
\newblock \emph{Nature}, 2026.
\newblock \doi{10.1038/s41586-026-11044-y}.
\newblock URL \url{https://doi.org/10.1038/s41586-026-11044-y}.
\newblock Published online 16 September 2026. Preprint: arXiv:2509.06917.

\bibitem[Wang et~al.(2026{\natexlab{b}})Wang, Cheng, Zuo, Zeng, He, Jiang,
  Yang, Wang, Zhao, Huang, Tian, Yuan, Zhong, Wang, Ding, Zhou, and
  Zhang]{wang2026naturebench}
Yuru Wang, Lejun Cheng, Yuxin Zuo, Sihang Zeng, Bingxiang He, Che Jiang, Junlin
  Yang, Yuchong Wang, Kaikai Zhao, Weifeng Huang, Kai Tian, Zhenzhao Yuan,
  Jincheng Zhong, Weizhi Wang, Ning Ding, Bowen Zhou, and Kaiyan Zhang.
\newblock {NatureBench}: Can coding agents match the published {SOTA} of
  {Nature}-family papers?
\newblock \emph{arXiv preprint arXiv:2606.24530}, 2026{\natexlab{b}}.
\newblock URL \url{https://arxiv.org/abs/2606.24530}.

\bibitem[Workman et~al.(2025)Workman, Yang, Muralidharan, and
  Le]{workman2025spatialbench}
Kenny Workman, Zhen Yang, Harihara Muralidharan, and Hannah Le.
\newblock {SpatialBench}: Can agents analyze real-world spatial biology data?
\newblock \emph{arXiv preprint arXiv:2512.21907}, 2025.
\newblock URL \url{https://arxiv.org/abs/2512.21907}.

\bibitem[St{\aa}hl et~al.(2016)St{\aa}hl, Salm{\'e}n, Vickovic, Lundmark,
  Navarro, Magnusson, Giacomello, Asp, Westholm, Huss, Mollbrink, Linnarsson,
  Codeluppi, Borg, Pont{\'e}n, Costea, Sahl{\'e}n, Mulder, Bergmann, Lundeberg,
  and Fris{\'e}n]{stahl2016st}
Patrik~L. St{\aa}hl, Fredrik Salm{\'e}n, Sanja Vickovic, Anna Lundmark,
  Jos{\'e}~Fern{\'a}ndez Navarro, Jens Magnusson, Stefania Giacomello, Michaela
  Asp, Jakub~O. Westholm, Mikael Huss, Annelie Mollbrink, Sten Linnarsson,
  Simone Codeluppi, {\AA}ke Borg, Fredrik Pont{\'e}n, Paul~Igor Costea, Pelin
  Sahl{\'e}n, Jan Mulder, Olaf Bergmann, Joakim Lundeberg, and Jonas
  Fris{\'e}n.
\newblock Visualization and analysis of gene expression in tissue sections by
  spatial transcriptomics.
\newblock \emph{Science}, 353\penalty0 (6294):\penalty0 78--82, 2016.
\newblock \doi{10.1126/science.aaf2403}.

\bibitem[Chen et~al.(2015)Chen, Boettiger, Moffitt, Wang, and
  Zhuang]{chen2015merfish}
Kok~Hao Chen, Alistair~N. Boettiger, Jeffrey~R. Moffitt, Siyuan Wang, and
  Xiaowei Zhuang.
\newblock Spatially resolved, highly multiplexed {RNA} profiling in single
  cells.
\newblock \emph{Science}, 348\penalty0 (6233):\penalty0 aaa6090, 2015.
\newblock \doi{10.1126/science.aaa6090}.

\bibitem[Janesick et~al.(2023)Janesick, Shelansky, Gottscho, Wagner, Williams,
  Rouault, Beliakoff, Morrison, Oliveira, Sicherman, Kohlway, Abousoud,
  Drennon, Mohabbat, {10x Development Teams}, and Taylor]{janesick2023xenium}
Amanda Janesick, Robert Shelansky, Andrew~D. Gottscho, Florian Wagner,
  Stephen~R. Williams, Morgane Rouault, Ghezal Beliakoff, Carolyn~A. Morrison,
  Michelli~F. Oliveira, Jordan~T. Sicherman, Andrew Kohlway, Jawad Abousoud,
  Tingsheng~Yu Drennon, Seayar~H. Mohabbat, {10x Development Teams}, and Sarah
  E.~B. Taylor.
\newblock High resolution mapping of the tumor microenvironment using
  integrated single-cell, spatial and in situ analysis.
\newblock \emph{Nature Communications}, 14\penalty0 (1):\penalty0 8353, 2023.
\newblock \doi{10.1038/s41467-023-43458-x}.

\bibitem[Zhang et~al.(2023{\natexlab{b}})Zhang, Deng, Kukanja, Agirre,
  Bartosovic, Dong, Ma, Ma, Su, Bao, Liu, Xiao, Rosoklija, Dwork, Mann, Leong,
  Boldrini, Wang, Haeussler, Raphael, Kluger, Castelo-Branco, and
  Fan]{zhang2023cuttag}
Di~Zhang, Yanxiang Deng, Petra Kukanja, Eneritz Agirre, Marek Bartosovic,
  Mingze Dong, Cong Ma, Sai Ma, Graham Su, Shuozhen Bao, Yang Liu, Yang Xiao,
  Gorazd~B. Rosoklija, Andrew~J. Dwork, J.~John Mann, Kam~W. Leong, Maura
  Boldrini, Liya Wang, Maximilian Haeussler, Benjamin~J. Raphael, Yuval Kluger,
  Gon{\char231}alo Castelo-Branco, and Rong Fan.
\newblock Spatial epigenome--transcriptome co-profiling of mammalian tissues.
\newblock \emph{Nature}, 616\penalty0 (7955):\penalty0 113--122,
  2023{\natexlab{b}}.
\newblock \doi{10.1038/s41586-023-05795-1}.

\bibitem[Wang et~al.(2022)Wang, Hu, Lv, Wang, Lan, Xiang, Tu, Wei, Han, Shi,
  Guo, Liu, Yang, Du, An, Cheng, Xu, Lu, Li, Zhang, Chen, Chen, Li, Wang, Xu,
  Hu, and Liu]{wang2022flysta3d}
Mingyue Wang, Qinan Hu, Tianhang Lv, Yuhang Wang, Qing Lan, Rong Xiang,
  Zhencheng Tu, Yanrong Wei, Kai Han, Chang Shi, Junfu Guo, Chao Liu, Tao Yang,
  Wensi Du, Yanru An, Mengnan Cheng, Jiangshan Xu, Haorong Lu, Wangsheng Li,
  Shaofang Zhang, Ao~Chen, Wei Chen, Yuxiang Li, Xiaoshan Wang, Xun Xu, Yuhui
  Hu, and Longqi Liu.
\newblock High-resolution {3D} spatiotemporal transcriptomic maps of developing
  {Drosophila} embryos and larvae.
\newblock \emph{Developmental Cell}, 57\penalty0 (10):\penalty0 1271--1283.e4,
  2022.
\newblock \doi{10.1016/j.devcel.2022.04.006}.

\bibitem[Jenkinson and Smith(2001)]{jenkinson2001flirt}
Mark Jenkinson and Stephen Smith.
\newblock A global optimisation method for robust affine registration of brain
  images.
\newblock \emph{Medical Image Analysis}, 5\penalty0 (2):\penalty0 143--156,
  2001.
\newblock \doi{10.1016/S1361-8415(01)00036-6}.

\bibitem[Avants et~al.(2008)Avants, Epstein, Grossman, and Gee]{avants2008syn}
Brian~B. Avants, Charles~L. Epstein, Murray Grossman, and James~C. Gee.
\newblock Symmetric diffeomorphic image registration with cross-correlation:
  evaluating automated labeling of elderly and neurodegenerative brain.
\newblock \emph{Medical Image Analysis}, 12\penalty0 (1):\penalty0 26--41,
  2008.
\newblock \doi{10.1016/j.media.2007.06.004}.

\bibitem[Sotiras et~al.(2013)Sotiras, Davatzikos, and
  Paragios]{sotiras2013registration}
Aristeidis Sotiras, Christos Davatzikos, and Nikos Paragios.
\newblock Deformable medical image registration: a survey.
\newblock \emph{IEEE Transactions on Medical Imaging}, 32\penalty0
  (7):\penalty0 1153--1190, 2013.
\newblock \doi{10.1109/TMI.2013.2265603}.

\bibitem[Zeira et~al.(2022)Zeira, Land, Strzalkowski, and
  Raphael]{zeira2022paste}
Ron Zeira, Max Land, Alexander Strzalkowski, and Benjamin~J. Raphael.
\newblock Alignment and integration of spatial transcriptomics data.
\newblock \emph{Nature Methods}, 19\penalty0 (5):\penalty0 567--575, 2022.
\newblock \doi{10.1038/s41592-022-01459-6}.

\bibitem[Liu et~al.(2023)Liu, Zeira, and Raphael]{liu2023paste2}
Xinhao Liu, Ron Zeira, and Benjamin~J. Raphael.
\newblock Partial alignment of multislice spatially resolved transcriptomics
  data.
\newblock \emph{Genome Research}, 33\penalty0 (7):\penalty0 1124--1132, 2023.
\newblock \doi{10.1101/gr.277670.123}.

\bibitem[Jones et~al.(2023)Jones, Townes, Li, and Engelhardt]{jones2023gpsa}
Andrew Jones, F.~William Townes, Didong Li, and Barbara~E. Engelhardt.
\newblock Alignment of spatial genomics data using deep {G}aussian processes.
\newblock \emph{Nature Methods}, 20\penalty0 (9):\penalty0 1379--1387, 2023.
\newblock \doi{10.1038/s41592-023-01972-2}.

\bibitem[Xia et~al.(2023)Xia, Cao, Tu, and Gao]{xia2023slat}
Chen-Rui Xia, Zhi-Jie Cao, Xin-Ming Tu, and Ge~Gao.
\newblock Spatial-linked alignment tool ({SLAT}) for aligning heterogenous
  slices.
\newblock \emph{Nature Communications}, 14\penalty0 (1):\penalty0 7236, 2023.
\newblock \doi{10.1038/s41467-023-43105-5}.

\bibitem[Clifton et~al.(2023)Clifton, Anant, Aihara, Atta, Aimiuwu, Kebschull,
  Miller, Tward, and Fan]{clifton2023stalign}
Kalen Clifton, Manjari Anant, Gohta Aihara, Lyla Atta, Osagie~K. Aimiuwu,
  Justus~M. Kebschull, Michael~I. Miller, Daniel Tward, and Jean Fan.
\newblock {STalign}: Alignment of spatial transcriptomics data using
  diffeomorphic metric mapping.
\newblock \emph{Nature Communications}, 14\penalty0 (1):\penalty0 8123, 2023.
\newblock \doi{10.1038/s41467-023-43915-7}.

\bibitem[Zhou et~al.(2023)Zhou, Dong, and Zhang]{zhou2023staligner}
Xiang Zhou, Kangning Dong, and Shihua Zhang.
\newblock Integrating spatial transcriptomics data across different conditions,
  technologies and developmental stages.
\newblock \emph{Nature Computational Science}, 3\penalty0 (10):\penalty0
  894--906, 2023.
\newblock \doi{10.1038/s43588-023-00528-w}.

\bibitem[Xu et~al.(2023)Xu, Wang, Fang, Luo, Chen, Wan, Wang, Tang, Xue, Li,
  Lin, and Qu]{xu2023spacel}
Hao Xu, Shuyan Wang, Minghao Fang, Songwen Luo, Chunpeng Chen, Siyuan Wan,
  Rirui Wang, Meifang Tang, Tian Xue, Bin Li, Jun Lin, and Kun Qu.
\newblock {SPACEL}: deep learning-based characterization of spatial
  transcriptome architectures.
\newblock \emph{Nature Communications}, 14\penalty0 (1):\penalty0 7603, 2023.
\newblock \doi{10.1038/s41467-023-43220-3}.

\bibitem[Qiu et~al.(2024)Qiu, Zhu, Lu, Yao, Jing, Min, Cheng, Pan, Zuo, King,
  Fang, Zheng, Wang, Wang, Zhang, Yu, Liao, Liu, Wu, Lai, Hao, Zhang, Wu,
  Zhang, Li, Tu, Lin, Yang, Li, Gu, Ellison, Lu, Hu, Hu, Chen, Liu, Weissman,
  Ma, Xu, Liu, and Bai]{qiu2024spateo}
Xiaojie Qiu, Daniel~Y. Zhu, Yifan Lu, Jiajun Yao, Zehua Jing, Kyung~Hoi Min,
  Mengnan Cheng, Hailin Pan, Lulu Zuo, Samuel King, Qi~Fang, Huiwen Zheng,
  Mingyue Wang, Shuai Wang, Qingquan Zhang, Sichao Yu, Sha Liao, Chao Liu,
  Xinchao Wu, Yiwei Lai, Shijie Hao, Zhewei Zhang, Liang Wu, Yong Zhang, Mei
  Li, Zhencheng Tu, Jinpei Lin, Zhuoxuan Yang, Yuxiang Li, Ying Gu, David
  Ellison, Yuancheng~Ryan Lu, Qinan Hu, Yuhui Hu, Ao~Chen, Longqi Liu,
  Jonathan~S. Weissman, Jiayi Ma, Xun Xu, Shiping Liu, and Yinqi Bai.
\newblock Spatiotemporal modeling of molecular holograms.
\newblock \emph{Cell}, 187\penalty0 (26):\penalty0 7351--7373.e61, 2024.
\newblock \doi{10.1016/j.cell.2024.10.011}.

\bibitem[Tang et~al.(2024)Tang, Luo, Zeng, Huang, Sui, Wu, and
  Wang]{tang2024cast}
Zefang Tang, Shuchen Luo, Hu~Zeng, Jiahao Huang, Xin Sui, Morgan Wu, and Xiao
  Wang.
\newblock Search and match across spatial omics samples at single-cell
  resolution.
\newblock \emph{Nature Methods}, 21\penalty0 (10):\penalty0 1818--1829, 2024.
\newblock \doi{10.1038/s41592-024-02410-7}.

\bibitem[Li et~al.(2024)Li, Lin, He, Han, Xu, Xu, Gao, Zhao, and
  Gao]{li2024santo}
Haoyang Li, Yingxin Lin, Wenjia He, Wenkai Han, Xiaopeng Xu, Chencheng Xu, Elva
  Gao, Hongyu Zhao, and Xin Gao.
\newblock {SANTO}: a coarse-to-fine alignment and stitching method for spatial
  omics.
\newblock \emph{Nature Communications}, 15\penalty0 (1):\penalty0 6048, 2024.
\newblock \doi{10.1038/s41467-024-50308-x}.

\bibitem[Klein et~al.(2025)Klein, Palla, Lange, Klein, Piran, Gander,
  Meng-Papaxanthos, Sterr, Saber, Jing, Bastidas-Ponce, Cota, Tarquis-Medina,
  Parikh, Gold, Lickert, Bakhti, Nitzan, Cuturi, and Theis]{klein2025moscot}
Dominik Klein, Giovanni Palla, Marius Lange, Michal Klein, Zoe Piran, Manuel
  Gander, Laetitia Meng-Papaxanthos, Michael Sterr, Lama Saber, Changying Jing,
  Aim{\'e}e Bastidas-Ponce, Perla Cota, Marta Tarquis-Medina, Shrey Parikh,
  Ilan Gold, Heiko Lickert, Mostafa Bakhti, Mor Nitzan, Marco Cuturi, and
  Fabian~J. Theis.
\newblock Mapping cells through time and space with {moscot}.
\newblock \emph{Nature}, 638\penalty0 (8052):\penalty0 1065--1075, 2025.
\newblock \doi{10.1038/s41586-024-08453-2}.

\bibitem[Yu and Xie(2025)]{yu2025stair}
Yuanyuan Yu and Zhi Xie.
\newblock Spatial transcriptomic alignment, integration, and {3D}
  reconstruction by {STAIR}.
\newblock \emph{Genome Biology}, 26\penalty0 (1):\penalty0 427, 2025.
\newblock \doi{10.1186/s13059-025-03895-x}.

\bibitem[Matthias et~al.(2026)Matthias, Yao, and Hicks]{albis}
Jan Matthias, Jianing Yao, and Stephanie~C. Hicks.
\newblock {ALBIS}: simulation of multi-resolution and multi-dimensional spatial
  transcriptomics data.
\newblock \emph{bioRxiv}, 2026.
\newblock \doi{10.64898/2026.10.06.757026}.
\newblock URL
  \url{https://www.biorxiv.org/content/10.64898/2026.10.06.757026v1}.

\bibitem[Kurtzer et~al.(2017)Kurtzer, Sochat, and
  Bauer]{kurtzer2017singularity}
Gregory~M. Kurtzer, Vanessa Sochat, and Michael~W. Bauer.
\newblock {Singularity}: Scientific containers for mobility of compute.
\newblock \emph{{PLOS ONE}}, 12\penalty0 (5):\penalty0 e0177459, 2017.
\newblock \doi{10.1371/journal.pone.0177459}.

\bibitem[Schott et~al.(2024)Schott, Le{\'o}n-Peri{\~n}{\'a}n, Splendiani,
  Strenger, Licha, Pentimalli, Schallenberg, Alles, Samut~Tagliaferro,
  Boltengagen, Ehrig, Abbiati, Dommerich, Pagani, Ferretti, Macino, Karaiskos,
  and Rajewsky]{schott2024openst}
Marie Schott, Daniel Le{\'o}n-Peri{\~n}{\'a}n, Elena Splendiani, Leon Strenger,
  Jan~Robin Licha, Tancredi~Massimo Pentimalli, Simon Schallenberg, Jonathan
  Alles, Sarah Samut~Tagliaferro, Anastasiya Boltengagen, Sebastian Ehrig,
  Stefano Abbiati, Steffen Dommerich, Massimiliano Pagani, Elisabetta Ferretti,
  Giuseppe Macino, Nikos Karaiskos, and Nikolaus Rajewsky.
\newblock {Open-ST}: high-resolution spatial transcriptomics in {3D}.
\newblock \emph{Cell}, 187\penalty0 (15):\penalty0 3953--3972.e26, 2024.
\newblock \doi{10.1016/j.cell.2024.05.055}.

\bibitem[Zeng et~al.(2023{\natexlab{a}})Zeng, Huang, Ren, Wang, Tang, Zhou,
  Zhou, Shi, Aditham, Sui, Chen, Lo, and Wang]{zeng2023ribomap}
Hu~Zeng, Jiahao Huang, Jingyi Ren, Connie~Kangni Wang, Zefang Tang, Haowen
  Zhou, Yiming Zhou, Hailing Shi, Abhishek Aditham, Xin Sui, Hongyu Chen,
  Jennifer~A. Lo, and Xiao Wang.
\newblock Spatially resolved single-cell translatomics at molecular resolution.
\newblock \emph{Science}, 380\penalty0 (6652):\penalty0 eadd3067,
  2023{\natexlab{a}}.
\newblock \doi{10.1126/science.add3067}.

\bibitem[Zeng et~al.(2023{\natexlab{b}})Zeng, Huang, Zhou, Meilandt, Dejanovic,
  Zhou, Bohlen, Lee, Ren, Liu, Tang, Sheng, Liu, Sheng, and
  Wang]{zeng2023starmapad}
Hu~Zeng, Jiahao Huang, Haowen Zhou, William~J. Meilandt, Borislav Dejanovic,
  Yiming Zhou, Christopher~J. Bohlen, Seung-Hye Lee, Jingyi Ren, Albert Liu,
  Zefang Tang, Hao Sheng, Jia Liu, Morgan Sheng, and Xiao Wang.
\newblock Integrative in situ mapping of single-cell transcriptional states and
  tissue histopathology in a mouse model of {Alzheimer's} disease.
\newblock \emph{Nature Neuroscience}, 26\penalty0 (3):\penalty0 430--446,
  2023{\natexlab{b}}.
\newblock \doi{10.1038/s41593-022-01251-x}.

\bibitem[Moffitt et~al.(2018)Moffitt, Bambah-Mukku, Eichhorn, Vaughn, Shekhar,
  Perez, Rubinstein, Hao, Regev, Dulac, and Zhuang]{moffitt2018mhpr}
Jeffrey~R. Moffitt, Dhananjay Bambah-Mukku, Stephen~W. Eichhorn, Eric Vaughn,
  Karthik Shekhar, Julio~D. Perez, Nimrod~D. Rubinstein, Junjie Hao, Aviv
  Regev, Catherine Dulac, and Xiaowei Zhuang.
\newblock Molecular, spatial, and functional single-cell profiling of the
  hypothalamic preoptic region.
\newblock \emph{Science}, 362\penalty0 (6416):\penalty0 eaau5324, 2018.
\newblock \doi{10.1126/science.aau5324}.

\bibitem[Ji et~al.(2020)Ji, Rubin, Thrane, Jiang, Reynolds, Meyers, Guo,
  George, Mollbrink, Bergenstr{\aa}hle, Larsson, Bai, Zhu, Bhaduri, Meyers,
  Rovira-Clav{\'e}, Hollmig, Aasi, Nolan, Lundeberg, and Khavari]{ji2020scc}
Andrew~L. Ji, Adam~J. Rubin, Kim Thrane, Sizun Jiang, David~L. Reynolds,
  Robin~M. Meyers, Margaret~G. Guo, Benson~M. George, Annelie Mollbrink, Joseph
  Bergenstr{\aa}hle, Ludvig Larsson, Yunhao Bai, Bokai Zhu, Aparna Bhaduri,
  Jordan~M. Meyers, Xavier Rovira-Clav{\'e}, S.~Tyler Hollmig, Sumaira~Z. Aasi,
  Garry~P. Nolan, Joakim Lundeberg, and Paul~A. Khavari.
\newblock Multimodal analysis of composition and spatial architecture in human
  squamous cell carcinoma.
\newblock \emph{Cell}, 182\penalty0 (2):\penalty0 497--514.e22, 2020.
\newblock \doi{10.1016/j.cell.2020.05.039}.

\bibitem[Sun et~al.(2021)Sun, Chen, Fischer, Lu, Zhan, Gillis, and
  Zador]{sun2021baristaseq}
Yu-Chi Sun, Xiaoyin Chen, Stephan Fischer, Shaina Lu, Huiqing Zhan, Jesse
  Gillis, and Anthony~M. Zador.
\newblock Integrating barcoded neuroanatomy with spatial transcriptional
  profiling enables identification of gene correlates of projections.
\newblock \emph{Nature Neuroscience}, 24\penalty0 (6):\penalty0 873--885, 2021.
\newblock \doi{10.1038/s41593-021-00842-4}.

\bibitem[Wang et~al.(2018)Wang, Allen, Wright, Sylwestrak, Samusik, Vesuna,
  Evans, Liu, Ramakrishnan, Liu, Nolan, Bava, and Deisseroth]{wang2018starmap}
Xiao Wang, William~E. Allen, Matthew~A. Wright, Emily~L. Sylwestrak, Nikolay
  Samusik, Sam Vesuna, Kathryn Evans, Cindy Liu, Charu Ramakrishnan, Jia Liu,
  Garry~P. Nolan, Felice-Alessio Bava, and Karl Deisseroth.
\newblock Three-dimensional intact-tissue sequencing of single-cell
  transcriptional states.
\newblock \emph{Science}, 361\penalty0 (6400):\penalty0 eaat5691, 2018.
\newblock \doi{10.1126/science.aat5691}.

\bibitem[Zhang et~al.(2021)Zhang, Eichhorn, Zingg, Yao, Cotter, Zeng, Dong, and
  Zhuang]{zhang2021mop}
Meng Zhang, Stephen~W. Eichhorn, Brian Zingg, Zizhen Yao, Kaelan Cotter,
  Hongkui Zeng, Hongwei Dong, and Xiaowei Zhuang.
\newblock Spatially resolved cell atlas of the mouse primary motor cortex by
  {MERFISH}.
\newblock \emph{Nature}, 598\penalty0 (7879):\penalty0 137--143, 2021.
\newblock \doi{10.1038/s41586-021-03705-x}.

\bibitem[Chen et~al.(2022)Chen, Liao, Cheng, Ma, Wu, Lai, Qiu, Yang, Xu, Hao,
  Wang, Lu, Chen, Liu, Huang, Li, Hong, Jiang, Peng, Liu, Shen, Liu, Li, Yuan,
  Wei, Zheng, Feng, Wang, Liu, Wang, et~al.]{chen2022stereoseq}
Ao~Chen, Sha Liao, Mengnan Cheng, Kailong Ma, Liang Wu, Yiwei Lai, Xiaojie Qiu,
  Jin Yang, Jiangshan Xu, Shijie Hao, Xin Wang, Huifang Lu, Xi~Chen, Xing Liu,
  Xin Huang, Zhao Li, Yan Hong, Yujia Jiang, Jian Peng, Shuai Liu, Mengzhe
  Shen, Chuanyu Liu, Quanshui Li, Yue Yuan, Xiaoyu Wei, Huiwen Zheng, Weimin
  Feng, Zhifeng Wang, Yang Liu, Zhaohui Wang, et~al.
\newblock Spatiotemporal transcriptomic atlas of mouse organogenesis using
  {DNA} nanoball-patterned arrays.
\newblock \emph{Cell}, 185\penalty0 (10):\penalty0 1777--1792.e21, 2022.
\newblock \doi{10.1016/j.cell.2022.04.003}.

\bibitem[Stickels et~al.(2021)Stickels, Murray, Kumar, Li, Marshall, Di~Bella,
  Arlotta, Macosko, and Chen]{stickels2021slideseqv2}
Robert~R. Stickels, Evan Murray, Pawan Kumar, Jilong Li, Jamie~L. Marshall,
  Daniela~J. Di~Bella, Paola Arlotta, Evan~Z. Macosko, and Fei Chen.
\newblock Highly sensitive spatial transcriptomics at near-cellular resolution
  with {Slide-seqV2}.
\newblock \emph{Nature Biotechnology}, 39\penalty0 (3):\penalty0 313--319,
  2021.
\newblock \doi{10.1038/s41587-020-0739-1}.

\bibitem[de~Oliveira et~al.(2025)de~Oliveira, Romero, Chung, Williams,
  Gottscho, Gupta, Pilipauskas, Mohabbat, Raman, Sukovich, Patterson, {Visium
  HD Development Team}, and Taylor]{oliveira2025visiumhd}
Michelli~Faria de~Oliveira, Juan~Pablo Romero, Meii Chung, Stephen~R. Williams,
  Andrew~D. Gottscho, Anushka Gupta, Susan~E. Pilipauskas, Seayar Mohabbat,
  Nandhini Raman, David~J. Sukovich, David~M. Patterson, {Visium HD Development
  Team}, and Sarah E.~B. Taylor.
\newblock High-definition spatial transcriptomic profiling of immune cell
  populations in colorectal cancer.
\newblock \emph{Nature Genetics}, 57\penalty0 (6):\penalty0 1512--1523, 2025.
\newblock \doi{10.1038/s41588-025-02193-3}.

\bibitem[Shi et~al.(2023)Shi, He, Zhou, Huang, Maher, Wang, Tang, Luo, Tan, Wu,
  Lin, Ren, Thapa, Tang, Chan, Deverman, Shen, Liu, Liu, and
  Wang]{shi2023starmapplus}
Hailing Shi, Yichun He, Yiming Zhou, Jiahao Huang, Kamal Maher, Brandon Wang,
  Zefang Tang, Shuchen Luo, Peng Tan, Morgan Wu, Zuwan Lin, Jingyi Ren, Yaman
  Thapa, Xin Tang, Ken~Y. Chan, Benjamin~E. Deverman, Hao Shen, Albert Liu, Jia
  Liu, and Xiao Wang.
\newblock Spatial atlas of the mouse central nervous system at molecular
  resolution.
\newblock \emph{Nature}, 622\penalty0 (7983):\penalty0 552--561, 2023.
\newblock \doi{10.1038/s41586-023-06569-5}.

\bibitem[He et~al.(2022)He, Bhatt, Brown, Brown, Buhr, Chantranuvatana,
  Danaher, Dunaway, Garrison, Geiss, Gregory, Hoang, Khafizov, Killingbeck,
  Kim, Kim, Kim, Klock, Korukonda, Kutchma, Lewis, Liang, Nelson, Ong, Perillo,
  Phan, Phan-Everson, Piazza, Rane, Reitz, et~al.]{he2022cosmx}
Shanshan He, Ruchir Bhatt, Carl Brown, Emily~A. Brown, Derek~L. Buhr, Kan
  Chantranuvatana, Patrick Danaher, Dwayne Dunaway, Ryan~G. Garrison, Gary
  Geiss, Mark~T. Gregory, Margaret~L. Hoang, Rustem Khafizov, Emily~E.
  Killingbeck, Dae Kim, Tae~Kyung Kim, Youngmi Kim, Andrew Klock, Mithra
  Korukonda, Alecksandr Kutchma, Zachary~R. Lewis, Yan Liang, Jeffrey~S.
  Nelson, Giang~T. Ong, Evan~P. Perillo, Joseph~C. Phan, Tien Phan-Everson,
  Erin Piazza, Tushar Rane, Zachary Reitz, et~al.
\newblock High-plex imaging of {RNA} and proteins at subcellular resolution in
  fixed tissue by spatial molecular imaging.
\newblock \emph{Nature Biotechnology}, 40\penalty0 (12):\penalty0 1794--1806,
  2022.
\newblock \doi{10.1038/s41587-022-01483-z}.

\end{thebibliography}
